\documentclass[twocolumn,linenumbers]{aastex63}
\usepackage{natbib}
\usepackage{gensymb}
\usepackage{amsmath}
\usepackage{enumitem}
\usepackage{graphicx}
\usepackage{multirow}
\usepackage[normalem]{ulem}
\usepackage{soul}
\usepackage{tabularx,booktabs}
\usepackage[caption=false]{subfig}
\usepackage{gensymb}
\usepackage{times}

\begin{document} 
\nolinenumbers
\title{\Large{Deep learning reconstruction of sunspot vector magnetic fields for forecasting solar storms}}

\correspondingauthor{Dattaraj B. Dhuri}
\email{dbd7602@nyu.edu}

\author{Dattaraj B. Dhuri}
\affiliation{Department of Astronomy and Astrophysics, Tata Institute of Fundamental Research, Mumbai, India 400005}
\affiliation{Center for Space Science, New York University Abu Dhabi, Abu Dhabi, UAE}

\author{Shamik Bhattacharjee}
\affiliation{Department of Astronomy and Astrophysics, Tata Institute of Fundamental Research, Mumbai, India 400005}

\author{Shravan M. Hanasoge}
\affiliation{Department of Astronomy and Astrophysics, Tata Institute of Fundamental Research, Mumbai, India 400005}
\affiliation{Center for Space Science, New York University Abu Dhabi, Abu Dhabi, UAE}

\author{Sashi Kiran Mahapatra}
\affiliation{Department of Astronomy and Astrophysics, Tata Institute of Fundamental Research, Mumbai, India 400005}

\begin{abstract}
\nolinenumbers
Solar magnetic activity produces extreme solar flares and coronal mass ejections, which pose grave threats to electronic infrastructure and can significantly disrupt economic activity. It is therefore important to appreciate the triggers of explosive solar activity and develop reliable space-weather forecasting. Photospheric vector-magnetic-field data capture sunspot magnetic-field complexity and can therefore improve the quality of space-weather prediction. However, state-of-the-art vector-field observations are consistently only available from Solar Dynamics Observatory/Helioseismic and Magnetic Imager (SDO/HMI) since 2010, with most other current and past missions and observational facilities such as Global Oscillations Network Group (GONG) only recording line-of-sight (LOS) fields.  Here, using an inception-based convolutional neural network, we reconstruct HMI sunspot vector-field features from LOS magnetograms of HMI as well as GONG with high fidelity ($\sim 90\%$ correlation) and sustained flare-forecasting accuracy. We rebuild vector-field features during the 2003 Halloween storms, for which only LOS-field observations are available, and the CNN-estimated electric-current-helicity accurately captures the observed rotation of the associated sunspot prior to the extreme flares, showing a striking increase. Our study thus paves the way for reconstructing three solar cycles worth of vector-field data from past LOS measurements, which are of great utility in improving space-weather forecasting models and gaining new insights about solar activity.
\end{abstract}

\section{Introduction}
Sunspot magnetic fields are generated within the solar interior, become buoyant through the solar convection zone and emerge at the photosphere and the corona as large-scale structures of sunspots and active regions (ARs) in the form of giant loops \citep{Cheung2014}. Coronal loops are dynamic, driven by emerging magnetic flux, electric current, and turbulent flows. Free magnetic energy stored in these loops is occasionally released via magnetic reconnection in the form of explosions such as flares and coronal mass ejections (CMEs) \citep{Shibata2011,Su2013}. Radiation and charged particles emitted in these explosions can lead to severe space weather, disrupting our life on Earth significantly \citep{Pulkkinen2005,Eastwood2017,Boteler2019}. In the past, the geomagnetic storm of 1989, resulting from a X15-class flare and subsequent CME, tripped circuit breakers in Hydro-Quebec power-grid causing a widespread blackout in Quebec \citep{Boteler2019}. The Halloween storm of 2003 produced extreme flares causing transformer malfunction and blackouts in Sweden, and damaging multiple science-mission satellites \citep{Pulkkinen2005}. In today's society, a high-magnitude solar storm can potentially lead to trillions of US dollars worth economic losses, with up to a decade of recovery time \citep{Eastwood2017}. Improving our understanding of AR magnetic-fields is therefore important for identifying triggers of these explosions and achieving reliable space-weather forecasting.

Coronal and photospheric AR magnetic fields are non-potential, comprising twisted flux-tubes as revealed by high-resolution, high-cadence observations of the SDO \citep{Pesnell-etall2012} since 2010. Large ARs and their complex dynamics, e.g. twisting and rotation, are known to be associated with solar explosive activity \citep{Toriumi2019}.  The SDO/HMI photospheric vector-magnetic-field observations facilitate the calculation of AR features \citep{Leka2008}, such as total unsigned magnetic flux, free energy density, electric current helicity and Lorentz forces, characterising the AR magnetic-field dynamics. These features are publicly available as the HMI data-product Space-weather HMI Active Region Patches (SHARPs) \citep{Bobra2014}. The SHARPs features are extensively used for statistical studies of pre-flare magnetic-field evolution and energy build up \citep{Dhuri2019} and improving space-weather forecasting using Machine Learning (ML) \citep{bobraflareprediction,Bobra2016,Chen2019}. HMI observations are limited to only one full solar cycle (cycle 24) and therefore, statistical space-weather forecasting models based on SHARPs are restricted.

Various difficulties are associated with the measurement of the transverse component of the photospheric magnetic-field \citep{Stenflo2013} and therefore,  ground- and space-based instruments monitoring the Sun since the 1970s, provide observations of only the longitudinal, i.e., line-of-sight (LOS) component. Continuous full-disk LOS field observations are available through the ground-based NASA/National Solar Observatory (NSO) Kitt Peak Telescope (1974 - present) \citep{Livingston:76}, space-based Michelson and Doppler Imager (MDI,1996 - 2011) \citep{Scherrer1995} and ground-based Global Oscillations Network Group (GONG,1995-present). These LOS-field measurements, although not sufficient for quantifying sunspot complexity to non-potential energy and helicity, have been useful for providing a qualitative assessment of AR morphology via sunspot classification schemes, such as the McIntosh classification \citep{McIntosh1990} and Mount Wilson classification which form the basis of operational space-weather forecasts \citep{crown2012validation}. 

Improving on these qualitative AR classifications and formally devising a method to quantify vector-field properties from LOS fields is of great utility --- (i) because it allows for ``improving" past datasets of LOS observations and understanding how vector-field features have evolved over multiple solar cycles, (ii) a reliable estimation of vector-field features over the past few decades can be used to build more robust statistical models for space-weather forecasting, and  (iii) for future missions acquiring only LOS data, vector-field features and even full vector-field construction can be part of an on-ground data-processing pipeline. ML methods such as convolutional neural networks (CNN) developed through the past decade have proven to be hugely successful in identifying patterns and correlations in large, high-dimensional datasets and particularly images \citep{LeCun2015, Goodfellow2016}. Here, we explore dependencies between LOS magnetograms and the corresponding full vector-field of ARs through a CNN model developed to estimate vector-field features SHARPs using the LOS magnetograms measurements from space-based SDO/HMI as well as ground-based GONG.

\section{Data \label{sec:Data}}
\begin{table}[t]
\centering
\begin{tabular}{lcc}
\toprule
 \multicolumn{3}{c}{Data} \\
\hline
      & Train \& Val & Test \\
      & May'10 - Sep'15 & Oct'15 - Aug'18 \\
\hline
    \# HMI ARs         &  848  & 194 \\
    \# HMI Samples         &  124633  & 26820\\
\hline
    \# GONG ARs         &  848  &  145 \\
    \# GONG Samples         &  114443  & 13454 \\
\bottomrule
\end{tabular}
\caption{The Helioseismic Magnetic Imager (HMI) and Global Oscillation Network Group (GONG) data used for training a CNN to obtain vector-field features from Space-Weather HMI Active Region Patches (SHARPs) \citep{Bobra2014}.}
\label{tab:data}
\end{table}
\begin{figure*}[t]
\centering
\includegraphics[width= \textwidth,trim={0.0cm 0.0cm 0.0cm 0.0cm},clip]{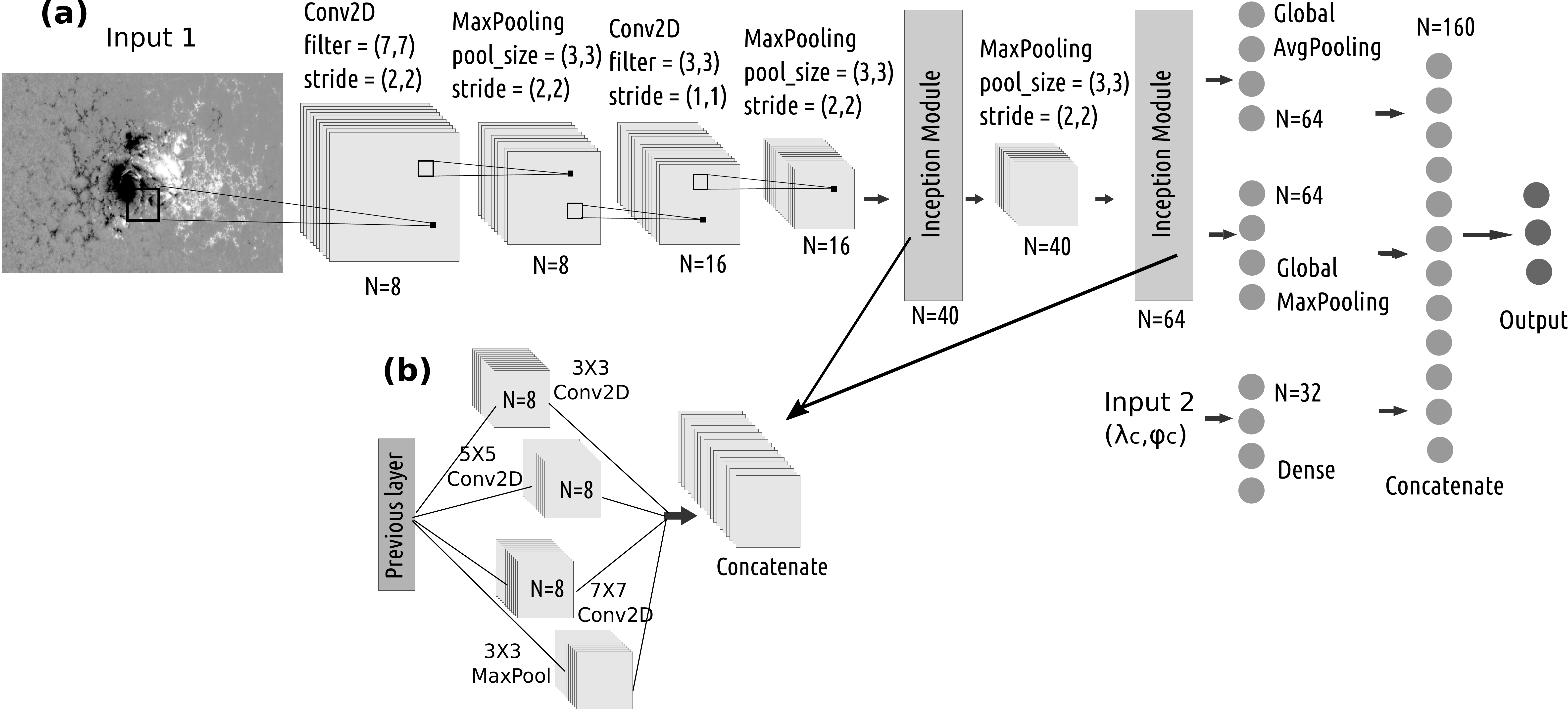}
\caption{{\bf The CNN architecture.} (a) Convolutional neural network (CNN) architecture used for obtaining vector-field features from LOS magnetograms. The architecture incorporates inception modules similar to GoogleNet \cite{GoogleNet2015}. The CNN takes in two inputs --- i) LOS magnetograms ii) AR center latitude ($\lambda_c$) and longitude ($\phi_c$). There are no fully connected layers that directly process the magnetogram input and therefore the CNN can process magnetogram patches of variable sizes. (b) Inception module used in the CNN.}
\label{fig:CNN}
\end{figure*}
We use photospheric LOS-magnetogram data provided by HMI and GONG. GONG provides only LOS magnetograms. HMI-derived SHARPs (the hmi.sharp\_cea\_720s data series \citep{Bobra2014}) include vector and LOS magnetograms of AR patches that are automatically detected and tracked as they rotate across the visible solar disk \citep{Bobra2014}. HMI magnetograms are available at a plate scale of $0.5\,{\rm arcsecs}$, i.e., $\sim 380\,{\rm km}$  at the disk center. GONG magnetograms are available at a plate scale of $2.5\,{\rm arcsecs}$. The magnetograms available in the hmi.sharp\_cea\_720s series are on a cylindrical equal-area (CEA) grid, thus eliminating the projection effects. We similarly remap the GONG AR magnetograms to a CEA grid. We train a CNN to obtain SHARPs features directly from LOS magnetograms of HMI as well as GONG. We only consider top SHARPs features that produce maximum flare forecasting accuracy for a machine learning (ML) model \citep{bobraflareprediction}. These are listed in Table~\ref{tab:dataNcorr}.

HMI measurements are sensitive to the observation conditions as well as the relative velocity between SDO and the Sun \citep{Hoeksema2014}. Observation conditions are indicated by the QUALITY flag and we consider measurements for which the Stokes vectors are reliable (QUALITY $\leq$ 10000 in hexadecimal) and when the relative velocity between SDO and the Sun is $< 3500\,{\rm m/s}$  \citep{Bobra2016}. Data closer to the limb are noisier because the higher relative velocities as well as projection effects. Therefore, we limit observations to within $\pm 45 \degree$ of the central meridian. Further, we only include ARs from the SHARPs data series that grow to a maximum area of $>25\,{\rm Mm}^2$. This eliminates a significant number of small ARs that do not produce major (M- or X-class) flares. The SHARPs feature calculation using HMI vector-field observations considers those pixels in the AR magnetograms for which the 180$\degree$ ambiguity resolution is reliable \citep{Bobra2014}.

Observations between May 2010 and Aug 2018 are used to train the CNN --- approximately $80\%$ of the data are used to train and validate the CNN, while the remaining is the unseen or test data. We chronologically split the ARs into training and validation and test partitions: ARs in the period May 2010 - Sep 2015 for training and validation and Oct 2015 - Aug 2018 is for the test. Six-hourly samples are drawn from the time series of each AR.  All samples from a given AR are exclusively part of either the training set, validation set or the test set, to avoid biases arising from temporal coherence of observations from an AR \citep{Ahmadzadeh_2021}. The number of ARs and magnetogram observations used for training, validation and test are listed in Table~\ref{tab:data}. Since solar activity depends on the phase of the cycle, the chronological splitting may introduce a bias for training the CNN. Indeed, the ratio of flaring-to-nonflaring ARs in the test data is approximately half its value in the training and validation dataset \citep{Bhattacharjee2020}. However, chronological splitting is appropriate for operational space-weather forecasting tools. 

\section{Methods \label{sec:Methods}}
\begin{figure*}[t]
\centering
\includegraphics[width= 0.8\textwidth,trim={0.0cm 0.0cm 0.0cm 0.0cm},clip]{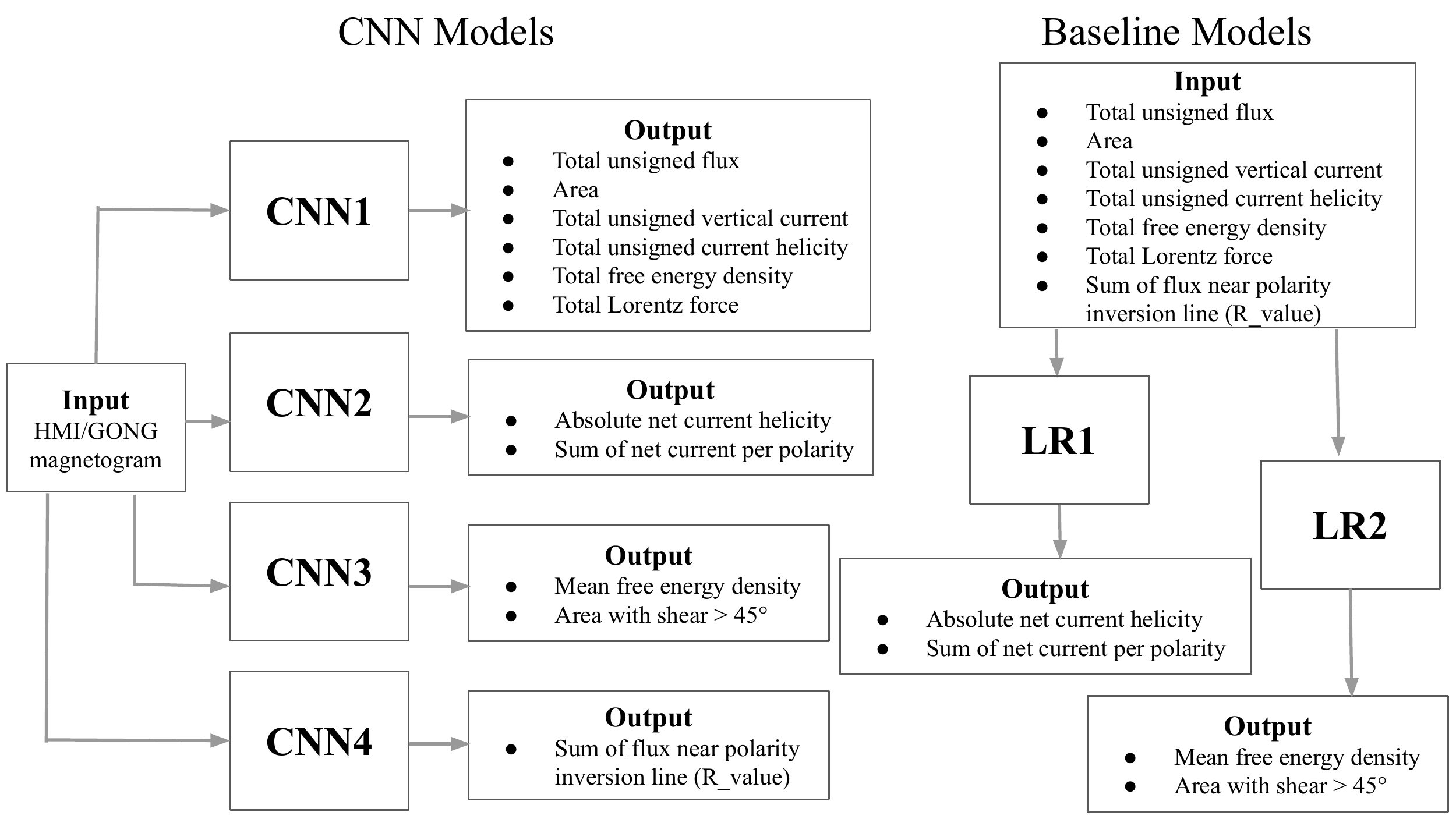}
\caption{{\bf Schematic of ML models.} ({\it left:}) Convolutional Neural Network (CNN) models process LOS magnetograms as input and produce vector-field features SHARPs as output. SHARPs features group together in four groups based on their mutual correlations as shown. We develop four different CNN models to estimate SHARPs from four different groups. All CNN models have identical architecture described in Figure~\ref{fig:CNN} except the final output layer, where, the number of output neurons is equal to the number of SHARPs features to be estimated from the respective group. ({\it right:}) Baseline models using Linear Regression (LR) for estimation of two groups of SHARPs features which depend on electric current and free energy respectively using extensive SHARPs features and Schrijver's R\_value \citep{Schrijver2007} as an input.}
\label{fig:CNNModels}
\end{figure*}
CNNs are neural networks with convolution filters (kernels) to scan over the input data, typically 2D data of images, and detect spatial patterns for tasks such as classification and identification \citep{LeCun2015,Goodfellow2016}. The  convolution filters are ${\rm K \times K}$ neurons that slide over the images and detect different patterns. Convolution filters have free parameters --- each neuron has weight $w$ and each convolution filter has bias $b$. Neurons process pixels of the inputs (or the outputs from previous layers) $\bf{x_i}$ by performing the operation $f\left(\sum_i{w_i x_i + b}\right)$, where $f$ is the activation function \citep{hastie01statisticallearning}.  CNNs also have pooling layers which are used to reduce the input size as it progresses to deeper levels of the CNN. A max-/average-pooling filter picks out the maximum or average value from the given ${\rm N \times N}$ feature map. Pooling layers typically follow a convolutional layer in a CNN to reduce the dimensionality. 

We use a CNN architecture with inception modules similar to inception V1 modules from GoogleNet \citep{GoogleNet2015}. Typically, in a convolutional layer, we use filters of fixed size that work best for the particular problem. However, inception modules are designed to detect patterns over a variety of length scales that may be present in the input. They involve convolution filters of different sizes in a single layer. The outputs from all the convolutional layers in an inception module are concatenated and supplied as an input to the following layer.  The inception module used here comprises three convolution filters of sizes $3\times3$, $5\times5$ and $7\times7$ and one $3\times3$ max-pooling filter.

The CNN architecture is shown in Figure~\ref{fig:CNN}. The CNN takes in two inputs --- i) LOS magnetograms of AR patches and ii) latitude ($\lambda_c$) and longitude ($\phi_c$) of the center of AR patches. The CNN consists of two regular convolutional layers and followed by the two inception modules that process the LOS magnetograms. The latitude and longitude are processed by a fully connected layer of neurons. The output of the two regular convolutional layers and the first inception module are reduced by a max-pooling layer. The output of the final inception module is reduced by a global max-pooling layer and also a global-average pooling layer. These are concatenated with the output from the fully connected layer that processes the longitude and latitude. The concatenated layer is connected to the output layer of neurons.  The number of neurons in the output layers is equal to the number of SHARPs features being estimated (see Figure~\ref{fig:CNNModels}). We use a {\it linear} activation function for the convolutional layers, which explicitly treats the positive and negative pixel values from LOS magnetograms symmetrically. Also, the fully connected layer of neurons has a {\it tanh} activation to explicitly treat positive and negative values of latitude and longitude, that are normalised between $\pm 1$, symmetrically. The final output layer of neurons have {\it sigmoid} activation \citep{10.1007/3-540-59497-3_175,hastie01statisticallearning} to yield the normalised value of the estimated SHARPs features between $0$ and $1$.

The absence of fully connected layers in the network that processes the LOS-magnetogram input implies that the CNN architecture can analyze LOS magnetograms of arbitrary sizes. Since AR patches are of varied dimensions, magnetograms in the training, validation and test data are also correspondingly differently sized. As such, our CNN does not require pre-processing to convert magnetograms to a fixed size and thus it is free from biases that may arise as a result of resizing \citep{Bhattacharjee2020}.

We use 10-times repeated-holdout validation for training the CNN \citep{hastie01statisticallearning}. We randomly split ARs in the training and validation sets into three parts and use data from two parts for training and the remaining part for validation. This process is repeated nine times while ensuring that the data from an AR is part of either the training or the validation and not both. The output from the CNN is compared to the original SHARPs feature values. The {\it sigmoid} output layer of the CNN lies in a continuous range between 0 to 1. The original SHARPs features are normalised by dividing by their respective maximum values. We partition the normalised features (over range 0 to 1) in the training set into ten bins of equal width (0.1) and oversample the data in each bin to match the number of samples in the maximally populated bin. The input magnetograms are standardised, i.e., a mean is subtracted and the resultant magnetogram is divided by a standard deviation of the magnetic field values. The mean and standard deviation used for standardisation are calculated over all pixels of all magnetograms in the training and validation data of the respective instrument. The CNN output is compared to the original SHARPs values and the loss function --- defined to be the mean squared error --- is computed. We train the CNN to minimize the mean squared error over different epochs using stochastic gradient descent \citep{Bottou91stochasticgradient,hastie01statisticallearning} with a learning rate of 0.00007. The CNN is developed using the Python library {\it keras}.  

\section{Results}
\begin{table*}[t]
\centering
\begin{tabular}{lcccc}
\hline
\hline
 &\multicolumn{2}{c}{10-times Repeated-Holdout Validation} & \multicolumn{2}{c}{Test}\\
SHARPs Features & HMI & GONG & HMI & GONG \\
\hline
Total unsigned flux               & 95.14 $\pm$ 00.62 & 90.87 $\pm$ 01.96 & 89.73 $\pm$ 02.70 &  87.42 $\pm$ 01.39 \\
Area                              & 95.87 $\pm$ 00.49 & 95.06 $\pm$ 00.84 & 92.00 $\pm$ 01.70 &  92.88 $\pm$ 00.88 \\ 
Total unsigned vertical current   & 94.78 $\pm$ 00.71 & 91.80 $\pm$ 01.69 & 88.86 $\pm$ 02.57 &  89.00 $\pm$ 01.76 \\
Total unsigned current helicity   & 95.74 $\pm$ 00.50 & 91.65 $\pm$ 01.76 & 88.33 $\pm$ 02.65 &  83.31 $\pm$ 02.28 \\
Total free energy density         & 96.19 $\pm$ 00.80 & 92.60 $\pm$ 01.60 & 90.17 $\pm$ 02.37 &  91.22 $\pm$ 01.25 \\ 
Total Lorentz force               & 96.64 $\pm$ 00.47 & 94.94 $\pm$ 00.98 & 90.63 $\pm$ 02.46 &  92.71 $\pm$ 00.87 \\
\hline
Absolute net current helicity     & 90.37 $\pm$ 03.28 & 63.76 $\pm$ 03.65 & 57.83 $\pm$ 08.84 &  57.60 $\pm$ 06.97 \\
Sum of net current per polarity   & 89.51 $\pm$ 02.53 & 64.58 $\pm$ 03.08 & 61.93 $\pm$ 07.63 &  59.09 $\pm$ 06.96 \\
\hline
Mean free energy density          & 95.10 $\pm$ 01.00 & 89.92 $\pm$ 00.79 & 92.13 $\pm$ 01.80 &  91.73 $\pm$ 00.54 \\ 
Area with shear $>45\degree$      & 95.02 $\pm$ 00.81 & 90.00 $\pm$ 01.19 & 90.59 $\pm$ 01.57 &  90.48 $\pm$ 00.46 \\ 
\hline
Flux near polarity inversion line & 90.54 $\pm$ 00.56 & 76.28 $\pm$ 01.83 & 77.11 $\pm$ 00.79 &  70.43 $\pm$ 00.79 \\ 
\hline
\hline
\end{tabular}
\caption{{\bf Pearson correlations  between the CNN-estimated vector-field features SHARPs and their true values.} SHARPs (Space-weather HMI Active Region Patches) features are calculated from HMI vector-field observations \citep{Bobra2014}. The p-values for all correlations are $\sim 0.0$. SHARPs features are mutually correlated \citep{Dhuri2019} and accordingly arranged in the four groups as features depending on (i) AR area (ii) electric current (iii) mean free energy density (iv) R-value i.e. the sum of flux near polarity inversion line \citep{Schrijver2007}.}
\label{tab:dataNcorr}
\end{table*}

\begin{table*}[t]
\centering
\begin{tabular}{lcccc}
\hline
\hline
 &\multicolumn{2}{c}{10-times Repeated-Holdout Validation} & \multicolumn{2}{c}{Test}\\
SHARPs Features & HMI & GONG & HMI & GONG \\
\hline
Total unsigned flux               & 86.92 $\pm$  01.13 & 86.27 $\pm$ 01.75 & 81.19 $\pm$  01.74 & 77.11 $\pm$  02.20\\ 
Area                              & 89.57 $\pm$  01.10 & 92.11 $\pm$ 00.75 & 87.10 $\pm$  01.33 & 86.91 $\pm$  01.21\\ 
Total unsigned vertical current   & 86.82 $\pm$  01.41 & 87.34 $\pm$ 01.38 & 81.22 $\pm$  02.10 & 79.07 $\pm$  02.52\\ 
Total unsigned current helicity   & 87.37 $\pm$  01.41 & 87.49 $\pm$ 01.48 & 82.61 $\pm$  02.24 & 79.40 $\pm$  02.69\\ 
Total free energy density         & 84.23 $\pm$  01.46 & 85.53 $\pm$ 02.17 & 81.32 $\pm$  03.27 & 80.11 $\pm$  02.42\\ 
Total Lorentz force               & 90.63 $\pm$  01.04 & 92.78 $\pm$ 00.98 & 86.96 $\pm$  01.51 & 86.49 $\pm$  01.66\\ 
\hline
Absolute net current helicity     & 59.61 $\pm$  05.26 & 59.35 $\pm$ 03.36 & 57.70 $\pm$  06.29 & 2.27 $\pm$  02.73\\ 
Sum of net current per polarity   & 60.02 $\pm$  03.95 & 65.75 $\pm$ 02.95 & 57.51 $\pm$  02.58 & 7.85 $\pm$  02.79\\ 
\hline
Mean free energy density          & 92.02 $\pm$  00.97 & 91.59 $\pm$ 00.92 & 93.02 $\pm$  01.42 & 92.58 $\pm$  00.35\\ 
Area with shear $>45\degree$      & 93.19 $\pm$  00.98 & 90.08 $\pm$ 00.54 & 91.63 $\pm$  00.65 & 89.96 $\pm$  00.37\\ 
\hline
Flux near polarity inversion line & 91.69 $\pm$  00.47 & 76.63 $\pm$ 02.60 & 83.00 $\pm$  00.85 & 75.32 $\pm$  00.92\\ 
\hline
\hline
\end{tabular}
\caption{{\bf Spearman correlations  between the CNN-estimated vector-field features SHARPs and their true values.} The p-values for all correlations are $\sim 0.0$.}
\label{tab:dataNcorrSp}
\end{table*}

\begin{table*}[t]
\centering
\begin{tabular}{lcccc}
\hline
\hline
 &\multicolumn{2}{c}{10-times Repeated-Holdout Validation} & \multicolumn{2}{c}{Test}\\
SHARPs Features & Pearson & Spearman & Pearson & Spearman \\
\hline
Absolute net current helicity     & $70.99 \pm 06.08$ & $68.21 \pm 02.25$ & $67.10 \pm 03.94$& $70.32 \pm 00.56$\\ 
Sum of net current per polarity   & $71.83 \pm 03.79$ & $70.51 \pm 02.23$ & $57.97 \pm 02.90$& $67.07 \pm 00.55$ \\ 
Mean free energy density          & $72.11 \pm 04.87$ & $79.63 \pm 04.49$ & $74.41 \pm 01.90$& $81.91 \pm 00.66$ \\ 
Area with shear $>45\degree$      & $67.89 \pm 02.93$ & $69.16 \pm 03.41$ & $70.83 \pm 00.39$& $77.86 \pm 00.20$ \\ 
\hline
\hline
\end{tabular}
\caption{{\bf Correlation between SHARPs features estimated using the baseline models and their true values.} Two baseline Linear Regression (LR) models are developed (Figure~\ref{fig:CNNModels}) that take extensive SHARPs features and Schrijver's R\_value as input and produce SHARPs features that depend on electric current and mean non-potential energy as output.The p-values for all correlations are $\sim 0.0$.}
\label{tab:baseline}
\end{table*}

\subsection{Estimation of AR vector-magnetic-field features using CNN}
\begin{figure*}[t]
\centering
\includegraphics[width= 0.85\textwidth,trim={1.5cm 2.4cm 1.3cm 2.3cm},clip]{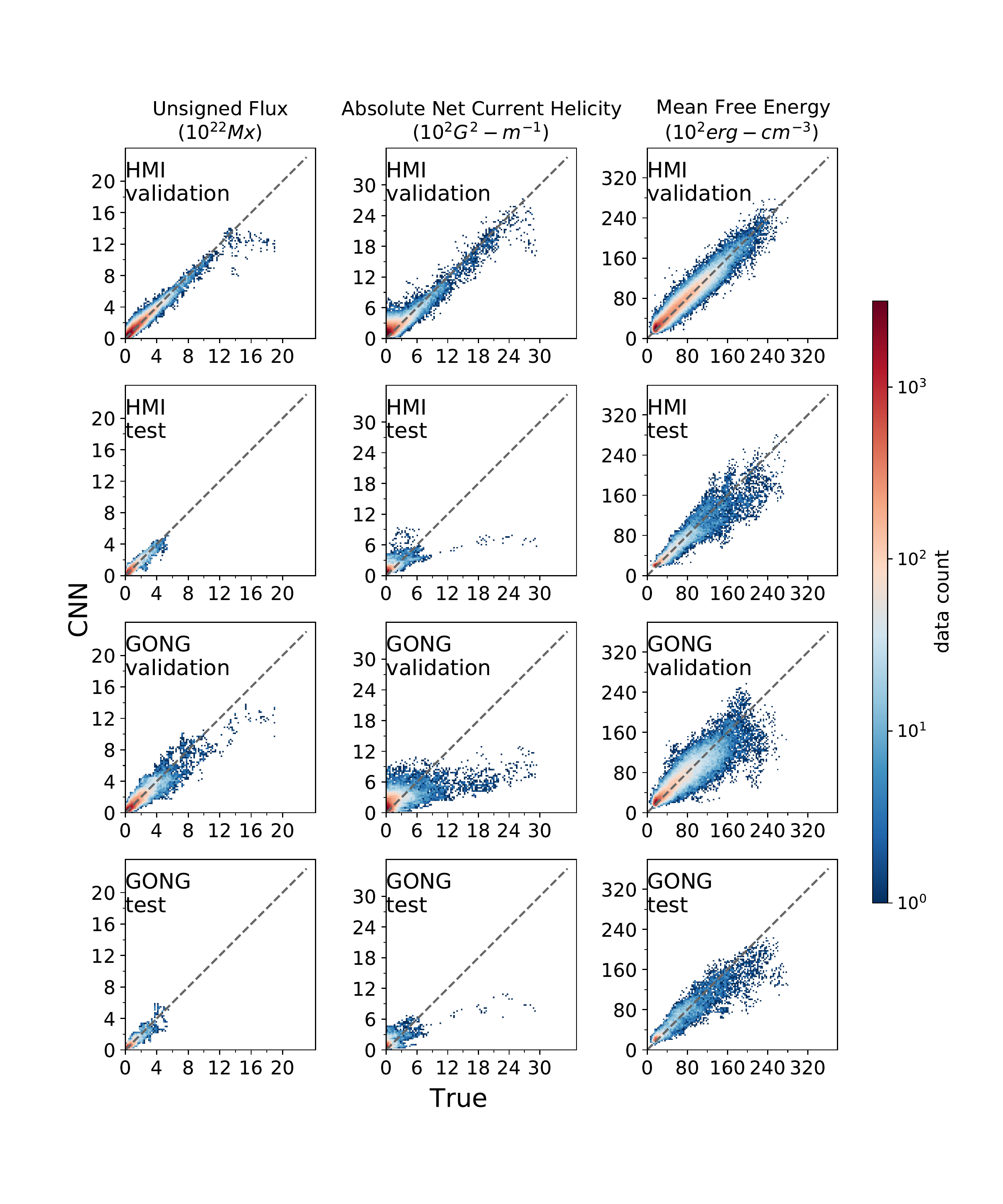}\\
\caption{{\bf Comparison of true and CNN-estimated SHARPs vector-field features.} Scatter plots for the distribution of CNN-estimated and true HMI values of total unsigned flux, absolute current helicity and mean free energy density obtained from HMI and GONG LOS magnetograms. The $45\degree$ line is marked for reference.}
\label{fig:corrNerror}
\end{figure*}
\begin{figure*}[t]
\centering
\subfloat{\includegraphics[width= 0.33\textwidth,trim={0.0cm 0.0cm 0.0cm 0.0cm},clip]{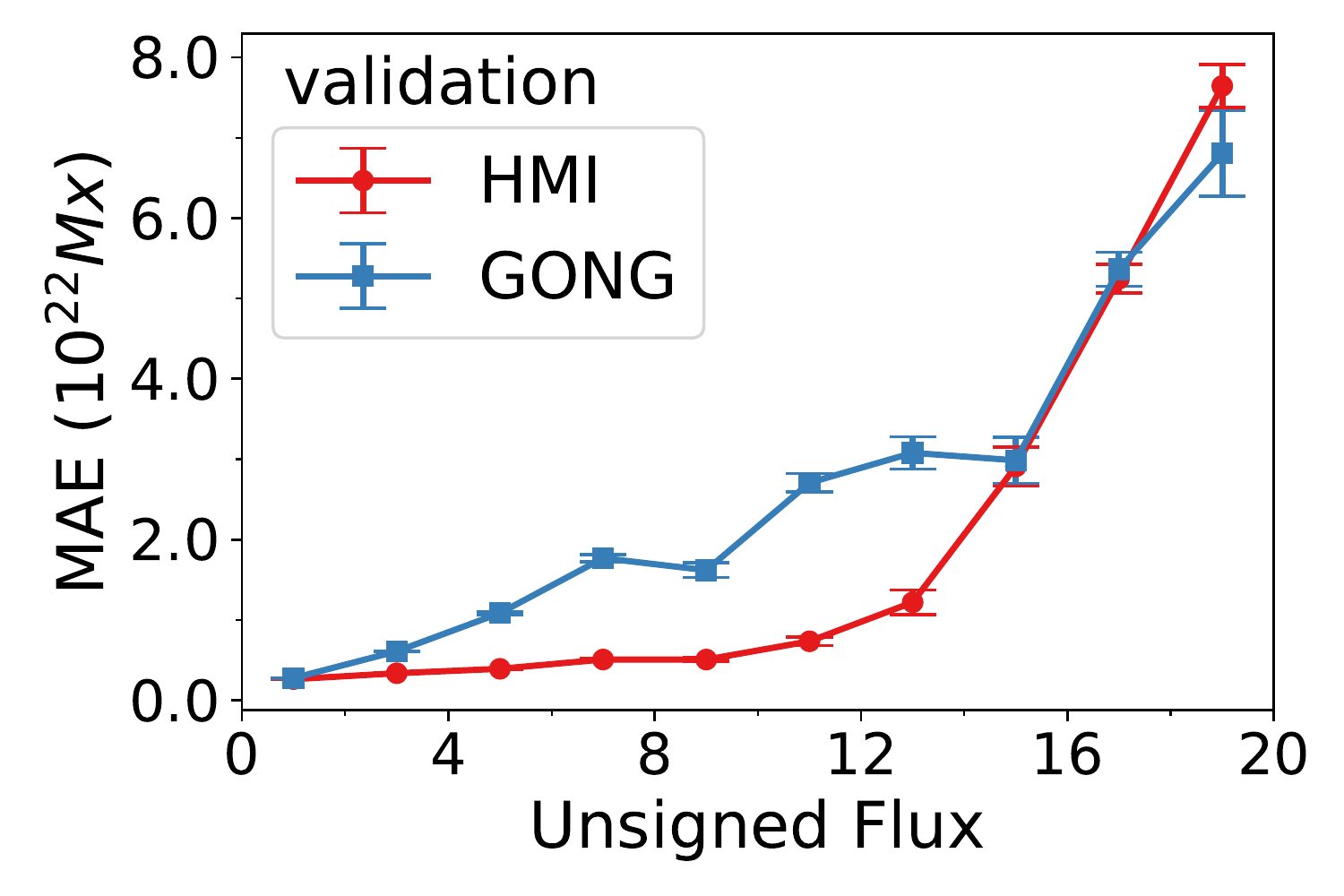}}
\subfloat{\includegraphics[width= 0.33\textwidth,trim={0.0cm 0.0cm 0.0cm 0.0cm},clip]{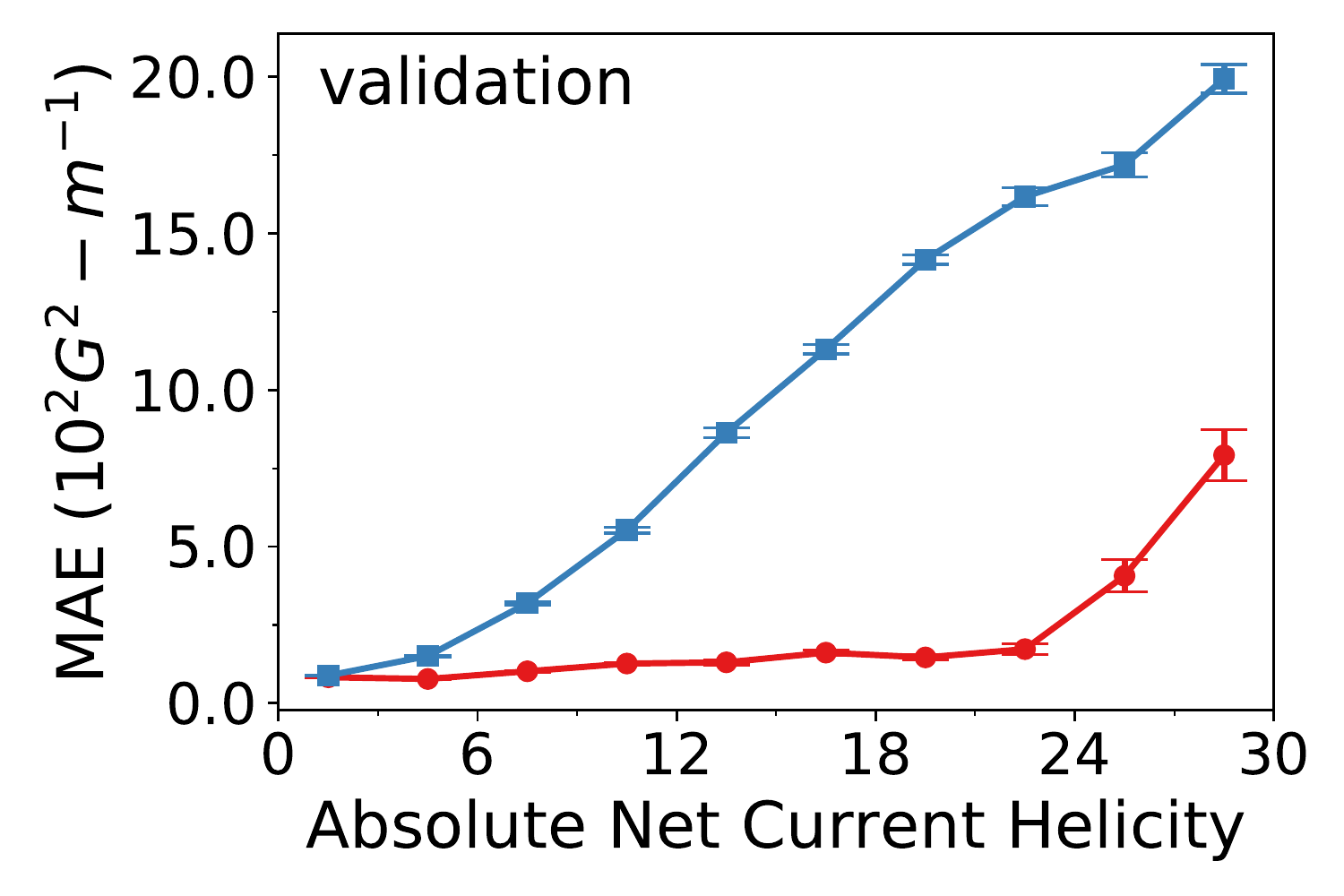}}
\subfloat{\includegraphics[width= 0.33\textwidth,trim={0.0cm 0.0cm 0.0cm 0.0cm},clip]{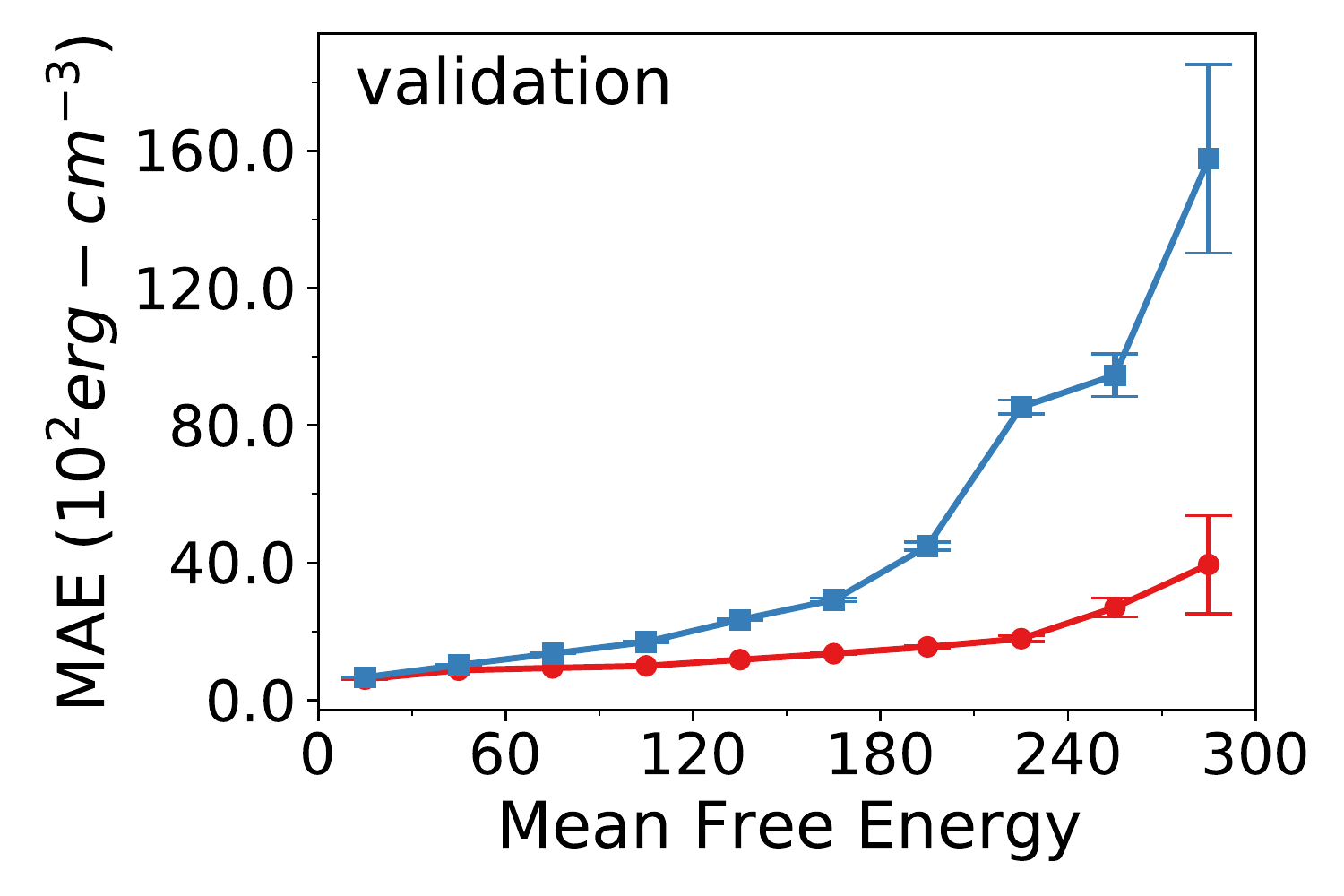}}\\
\subfloat{\includegraphics[width= 0.33\textwidth,trim={0.0cm 0.0cm 0.0cm 0.0cm},clip]{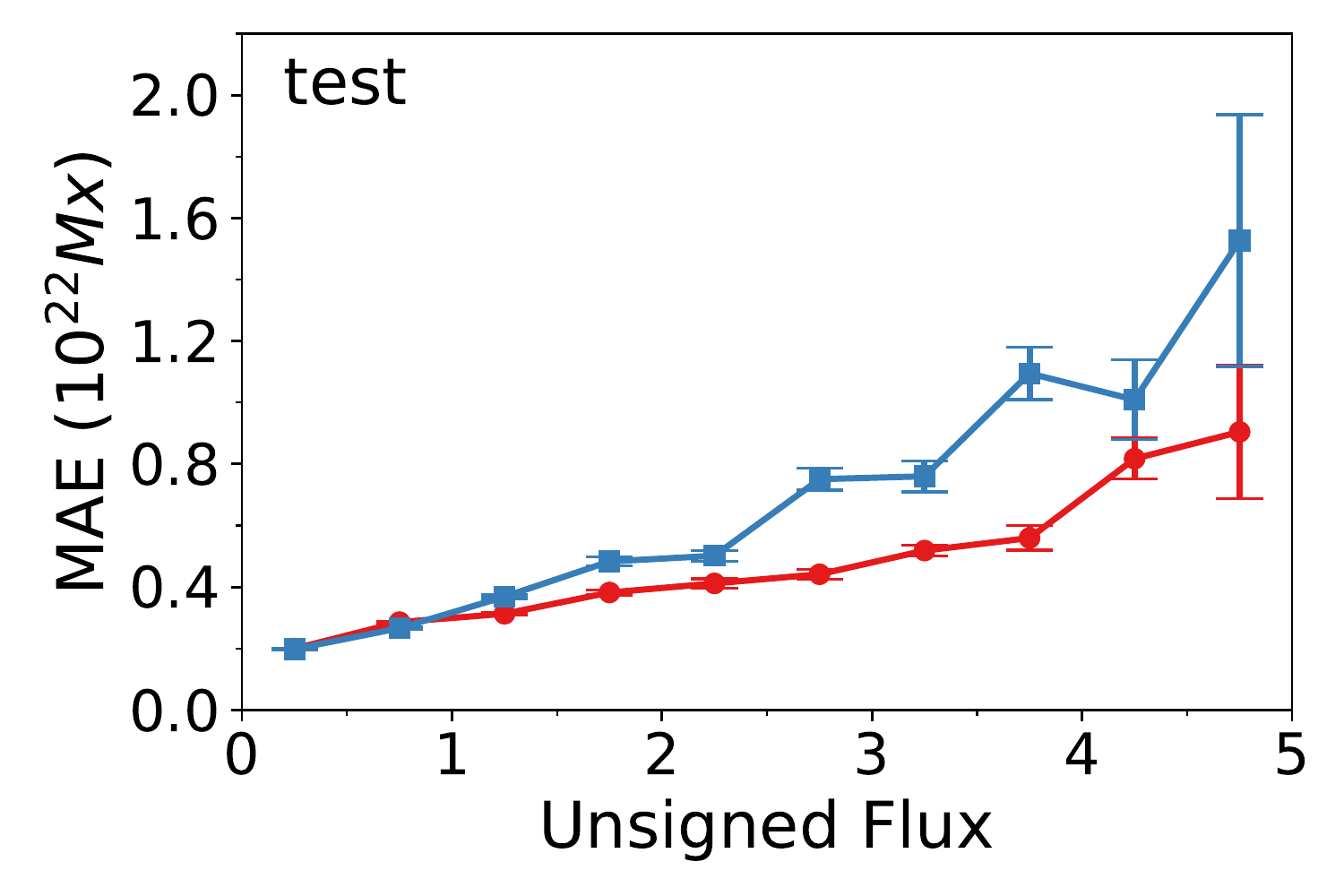}}
\subfloat{\includegraphics[width= 0.33\textwidth,trim={0.0cm 0.0cm 0.0cm 0.0cm},clip]{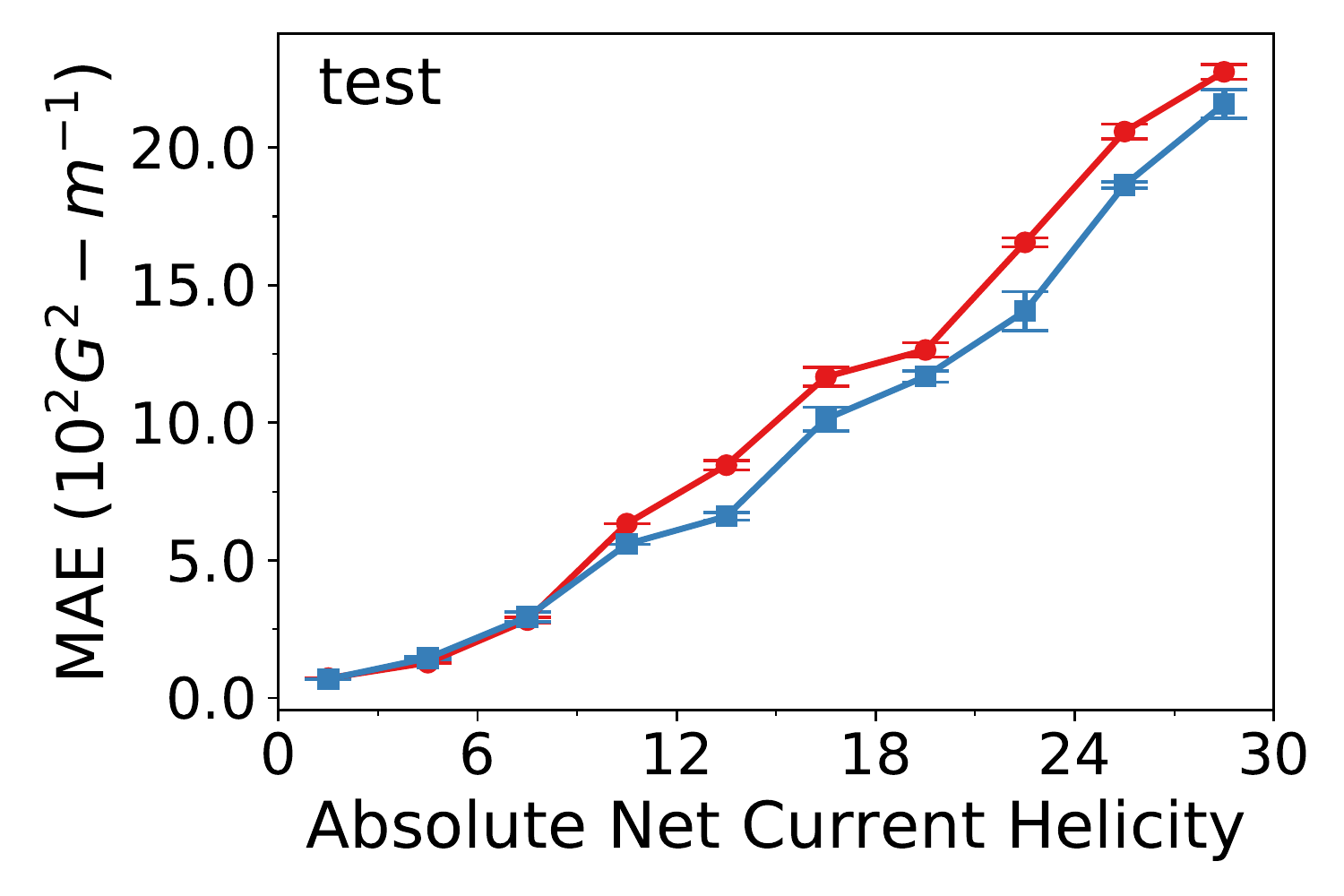}}
\subfloat{\includegraphics[width= 0.33\textwidth,trim={0.0cm 0.0cm 0.0cm 0.0cm},clip]{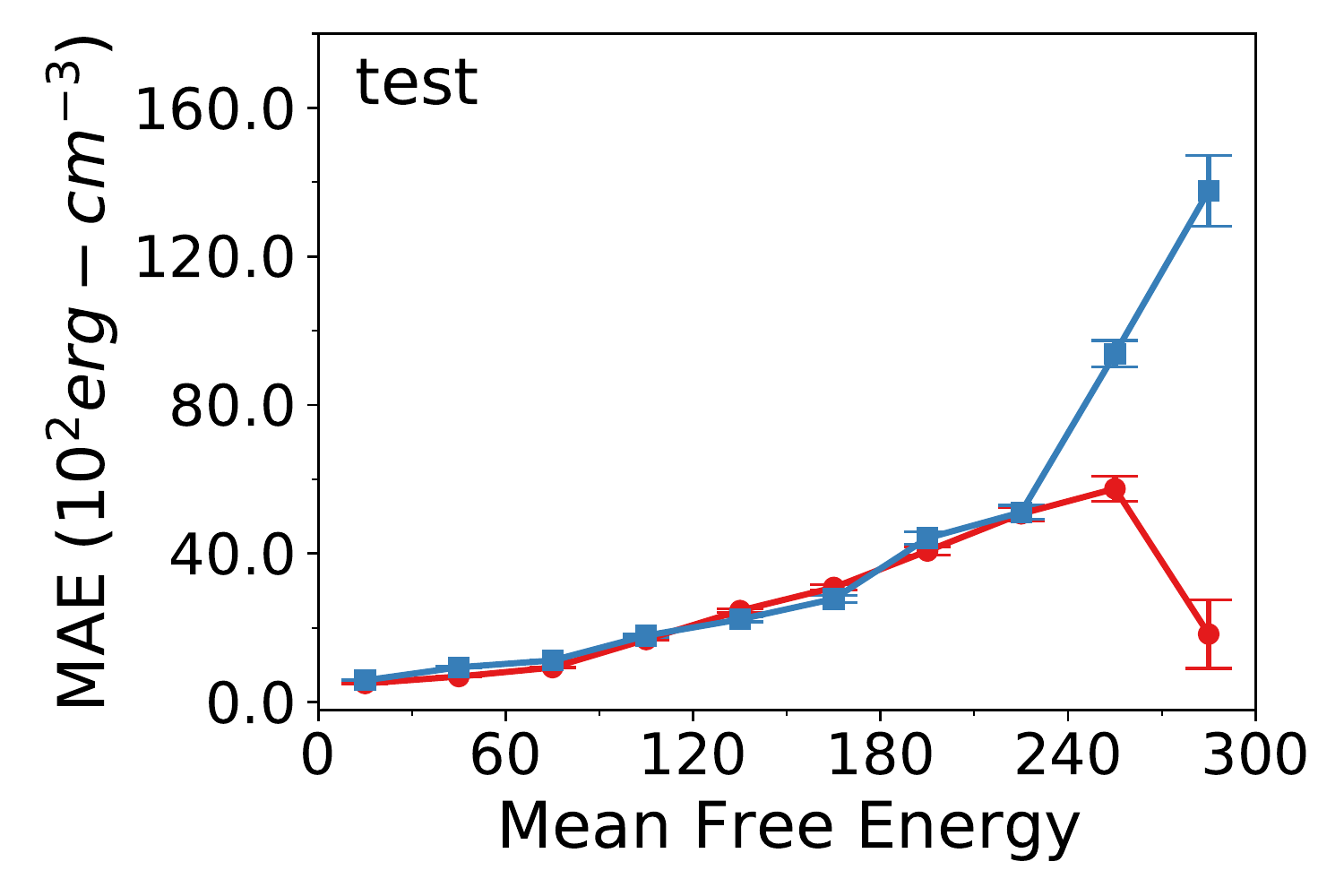}}
\caption{ {\bf Mean absolute error of the CNN-reconstructed vector-field features.} The CNN-estimated values of total unsigned flux, absolute current helicity and mean free-energy density for HMI and GONG are binned into 10 uniform bins as per the respective true values. Mean absolute error as a function of the mean true values of each population bin are shown. 1-$\sigma$ error bars for each bin are also shown. The legend in the top left panel applies to all panels.}
\label{fig:mae}
\end{figure*}

\begin{figure*}[t]
\centering
\includegraphics[width= 0.75\textwidth,trim={0.50cm 0.60cm 0.80cm 0.4cm},clip]{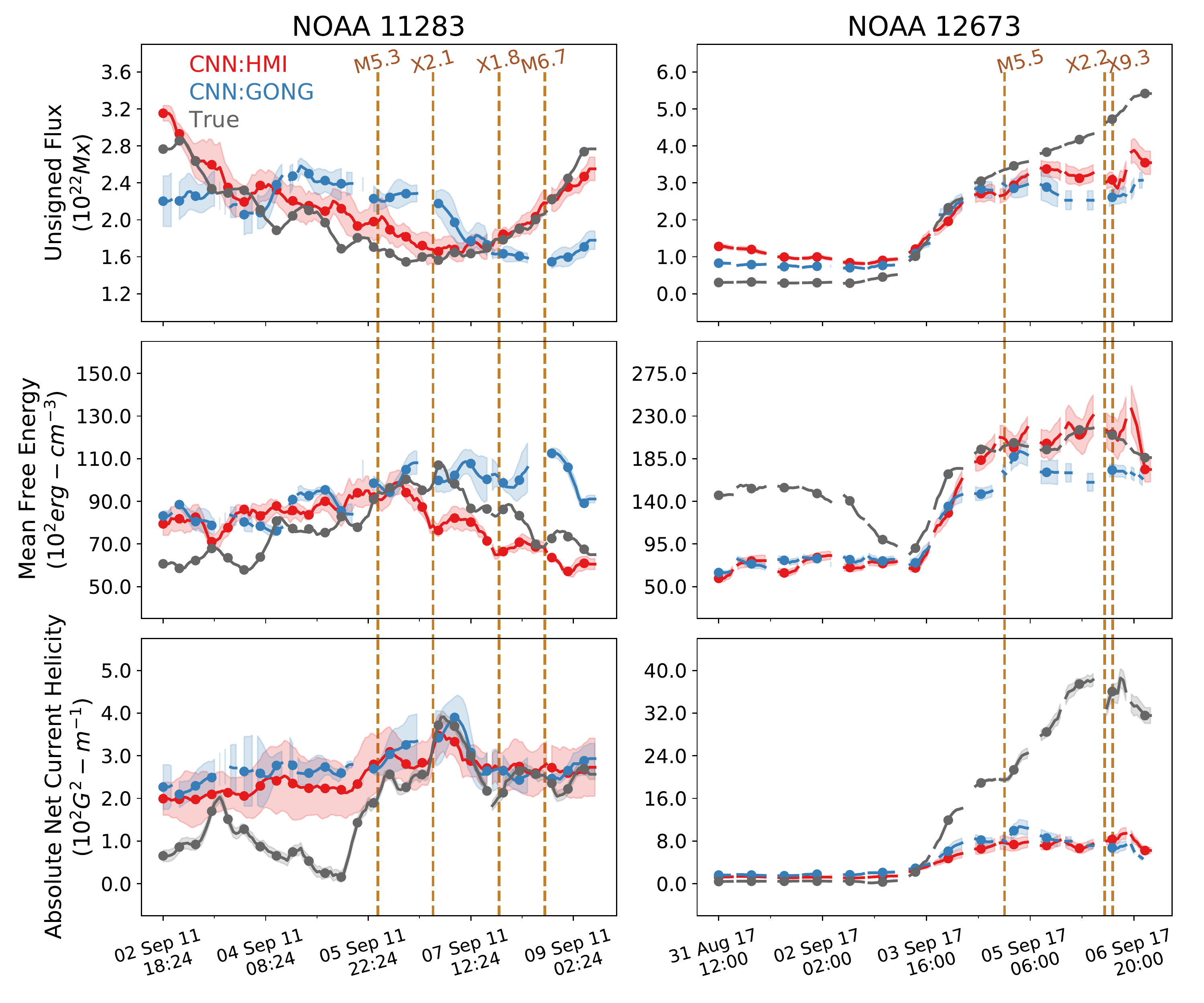}\\
\caption{{\bf Comparison of true and CNN-estimated SHARPs vector-field features.} Comparisons of time evolution of the CNN-estimated total unsigned flux, absolute net current helicity and mean free energy density with true values for ARs that produce M5 or greater flares. Only AR observations within $\pm 45\degree$ of the central meridian are considered. The left plot shows a typical result (see Appendix~Figure~\ref{fig:allTE} for all ARs with major flares). The right plot shows an extreme event with the largest flare observed in cycle 24. The gaps correspond to the missing observations and 1-$\sigma$ error bars are shown. The legend in top left applies to all plots. The plots are smoothed with a six hourly running average.}
\label{fig:TE}
\end{figure*}
\begin{figure*}
\centering
\includegraphics[width= 0.8\textwidth,trim={0.3cm 0.4cm 0.3cm 0.1cm},clip]{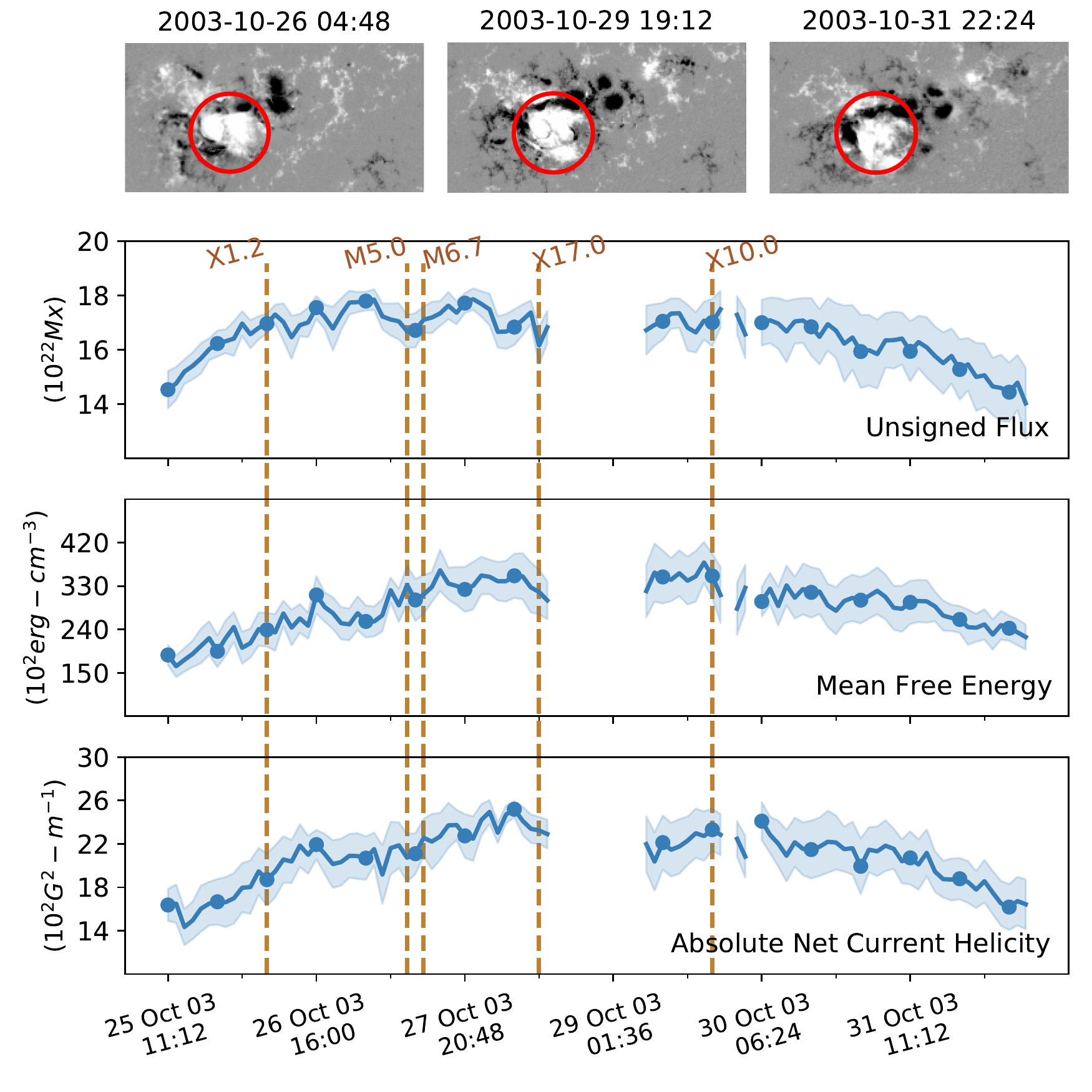}\\
\caption{{\bf The 2003 Halloween Storms}. {\it(top)} AR NOAA 10486 MDI magnetograms show the AR dynamics leading to extreme flare events X17.0, X10.0 and X28.0 (outside $\pm 45\degree$ of the central meridian, not shown) characterised by rotation of the positive polarity spot \citep{Zhang2008,Kazachenko_2010}. highlighted within the red circle. Positive (white) and negative (black) polarities are saturated at 1000~G. {\it(bottom)} The CNN-estimated total unsigned flux, mean-free-energy density and absolute net current helicity during these storms showing a systematic rise leading to the flares. The CNN accurately captures helicity injection due to the sunspot rotation showing a significant increase in the mean-free-energy density and electric current helicity. The gaps correspond to the missing observations and 1-$\sigma$ error bars are shown. The plots are smoothed with a six hourly running average.} 
\label{fig:timeEvoleHalloween}
\end{figure*}

\begin{table*}[t]
\centering
\begin{tabular}{lccccccc}
\hline
\hline
 & &\multicolumn{2}{c}{Total unsigned flux} & \multicolumn{2}{c}{Absolute net current helicity} & \multicolumn{2}{c}{Mean free energy density}\\
& & Spline Fit & Time Derivative & Spline Fit & Time Derivative & Spline Fit & Time Derivative \\
\hline
Validation      & HMI   & 97.41 $\pm$ 00.37 & 84.46 $\pm$ 02.75 & 94.99 $\pm$ 02.02 & 77.70 $\pm$ 12.16 & 98.54 $\pm$ 00.38 & 89.68 $\pm$ 02.50\\
                      & GONG  & 91.62 $\pm$ 01.94 & 25.51 $\pm$ 08.68 & 66.12 $\pm$ 03.73 & 33.23 $\pm$ 05.40 & 92.12 $\pm$ 07.77 & 75.52 $\pm$ 02.29 \\
\hline
\multirow{2}{*}{Test} & HMI   & 90.50 $\pm$ 02.60 & 75.18 $\pm$ 06.81 & 60.21 $\pm$ 09.21 & 32.44 $\pm$ 18.33 & 94.14 $\pm$ 01.52 & 82.06 $\pm$ 03.06 \\
                      & GONG  & 88.63 $\pm$ 01.48 & 18.07 $\pm$ 06.22 & 59.72 $\pm$ 07.14 & 27.03 $\pm$ 07.81 & 93.69 $\pm$ 00.49 & 70.52 $\pm$ 01.91 \\
\hline
\end{tabular}
\caption{{\bf  Trend comparison of CNN-estimated and true SHARPs features.} Pearson correlations between the time derivative of the true and the CNN-estimated values of total unsigned flux, absolute net current helicity and mean free energy density. The time derivative is obtained for the true and the CNN-estimated features of each AR after fitting the respective time series to a cubic spline. For reference, the Pearson correlations between the spline fit values of the true and the CNN-estimated features are also shown. Note that these are consistent with the values in Table~\ref{tab:dataNcorr}.}
\label{tab:Trend_Measures}
\end{table*}
\begin{table*}
\centering
\begin{tabular}{lccccc}
\hline
\hline
 Feature & accuracy & recall(+) & recall(-) & TSS \\
\hline
\hline
 \multicolumn{5}{c}{True SHARPs} \\
\hline
Total unsigned flux                      & 87.04 $\pm$ 02.36 & 51.16 $\pm$ 18.79 & 88.99 $\pm$ 02.95 & 40.16 $\pm$ 17.15 \\ 
Area                                     & 85.27 $\pm$ 02.16 & 58.78 $\pm$ 17.32 & 86.67 $\pm$ 02.71 & 45.45 $\pm$ 15.55 \\ 
Total unsigned vertical current          & 88.19 $\pm$ 02.75 & 64.39 $\pm$ 07.67 & 89.50 $\pm$ 03.00 & 53.89 $\pm$ 07.14 \\ 
Total unsigned current helicity          & 89.81 $\pm$ 02.82 & 66.17 $\pm$ 05.76 & 91.15 $\pm$ 02.99 & 57.32 $\pm$ 06.20 \\ 
Total free energy density                & 89.77 $\pm$ 02.04 & 52.17 $\pm$ 11.96 & 91.84 $\pm$ 02.33 & 44.01 $\pm$ 11.07 \\ 
Total Lorentz force                      & 87.68 $\pm$ 02.47 & 52.39 $\pm$ 16.75 & 89.60 $\pm$ 02.90 & 41.99 $\pm$ 15.55 \\ 
Absolute net current helicity            & 91.01 $\pm$ 02.17 & 58.29 $\pm$ 08.63 & 92.92 $\pm$ 02.40 & 51.21 $\pm$ 08.32 \\ 
Sum of net current per polarity          & 90.96 $\pm$ 01.99 & 60.64 $\pm$ 08.29 & 92.72 $\pm$ 02.30 & 53.37 $\pm$ 07.56 \\ 
Mean free energy density                 & 76.56 $\pm$ 02.94 & 70.16 $\pm$ 04.95 & 76.92 $\pm$ 03.09 & 47.08 $\pm$ 05.98 \\ 
Area with shear $>45\degree$             & 67.15 $\pm$ 03.32 & 74.55 $\pm$ 04.39 & 66.72 $\pm$ 03.54 & 41.27 $\pm$ 05.54 \\ 
Log of flux near polarity inversion line & 67.57 $\pm$ 03.83 & 97.98 $\pm$ 01.63 & 65.83 $\pm$ 04.05 & 63.81 $\pm$ 05.06 \\ 
\hline
 \multicolumn{5}{c}{CNN:HMI}\\
\hline
Total unsigned flux                      & 85.01 $\pm$ 02.69 & 69.52 $\pm$ 08.66 & 85.87 $\pm$ 02.99 & 55.39 $\pm$ 07.65 \\ 
Area                                     & 83.54 $\pm$ 01.84 & 70.88 $\pm$ 06.98 & 84.23 $\pm$ 02.17 & 55.11 $\pm$ 05.36 \\ 
Total unsigned vertical current          & 85.51 $\pm$ 02.59 & 71.04 $\pm$ 07.54 & 86.30 $\pm$ 02.94 & 57.34 $\pm$ 06.02 \\ 
Total unsigned current helicity          & 86.39 $\pm$ 02.23 & 69.81 $\pm$ 06.28 & 87.32 $\pm$ 02.46 & 57.13 $\pm$ 05.36 \\ 
Total free energy density                & 88.02 $\pm$ 02.00 & 61.87 $\pm$ 08.08 & 89.48 $\pm$ 02.30 & 51.35 $\pm$ 06.68 \\ 
Total Lorentz force                      & 86.04 $\pm$ 02.74 & 68.47 $\pm$ 09.26 & 87.03 $\pm$ 02.94 & 55.50 $\pm$ 08.97 \\ 
Absolute net current helicity            & 88.70 $\pm$ 02.03 & 72.11 $\pm$ 06.77 & 89.62 $\pm$ 02.14 & 61.73 $\pm$ 06.52 \\ 
Sum of net current per polarity          & 88.82 $\pm$ 02.22 & 73.94 $\pm$ 08.13 & 89.64 $\pm$ 02.42 & 63.58 $\pm$ 07.50 \\ 
Mean free energy density                 & 76.33 $\pm$ 02.68 & 69.17 $\pm$ 05.34 & 76.72 $\pm$ 02.87 & 45.89 $\pm$ 05.40 \\ 
Area with shear $>45\degree$             & 71.55 $\pm$ 02.72 & 70.73 $\pm$ 08.44 & 71.57 $\pm$ 03.00 & 42.30 $\pm$ 07.84 \\ 
Log of flux near polarity inversion line & 76.62 $\pm$ 02.80 & 91.88 $\pm$ 03.58 & 75.72 $\pm$ 03.16 & 67.59 $\pm$ 02.28 \\ 
\hline
 \multicolumn{5}{c}{CNN:GONG}\\
\hline
Total unsigned flux                      & 87.01 $\pm$ 02.44 & 53.17 $\pm$ 11.56 & 88.87 $\pm$ 02.84 & 42.05 $\pm$ 10.34 \\ 
Area                                     & 84.32 $\pm$ 02.57 & 60.14 $\pm$ 12.87 & 85.62 $\pm$ 02.98 & 45.76 $\pm$ 11.51 \\ 
Total unsigned vertical current          & 87.06 $\pm$ 02.43 & 54.33 $\pm$ 11.78 & 88.87 $\pm$ 02.83 & 43.20 $\pm$ 10.66 \\ 
Total unsigned current helicity          & 88.15 $\pm$ 02.42 & 49.19 $\pm$ 13.78 & 90.30 $\pm$ 02.83 & 39.49 $\pm$ 12.85 \\ 
Total free energy density                & 89.44 $\pm$ 01.69 & 41.35 $\pm$ 17.17 & 92.11 $\pm$ 02.34 & 33.45 $\pm$ 15.85 \\ 
Total Lorentz force                      & 86.96 $\pm$ 02.39 & 48.25 $\pm$ 19.33 & 89.06 $\pm$ 02.82 & 37.30 $\pm$ 18.34 \\ 
Absolute net current helicity            & 91.26 $\pm$ 02.18 & 54.50 $\pm$ 11.18 & 93.39 $\pm$ 02.34 & 47.89 $\pm$ 11.16 \\ 
Sum of net current per polarity          & 90.79 $\pm$ 01.77 & 52.30 $\pm$ 10.90 & 93.02 $\pm$ 02.06 & 45.32 $\pm$ 10.61 \\ 
Mean free energy density                 & 77.19 $\pm$ 03.73 & 67.10 $\pm$ 07.69 & 77.76 $\pm$ 04.08 & 44.86 $\pm$ 07.50 \\ 
Area with shear $>45\degree$             & 70.50 $\pm$ 04.47 & 73.21 $\pm$ 07.29 & 70.32 $\pm$ 04.84 & 43.53 $\pm$ 07.77 \\ 
Log of flux near polarity inversion line & 78.22 $\pm$ 04.06 & 84.98 $\pm$ 03.89 & 77.82 $\pm$ 04.38 & 62.80 $\pm$ 04.83 \\ 
\hline
\hline
\end{tabular}
\caption{{\bf A comparison of the CNN-estimated and true SHARPs features for flare forecasting using linear discriminant analysis (LDA) of each feature.} 1-$\sigma$ standard deviation is shown.}
\label{tab:LDA}
\end{table*}

\begin{table*}[t]
\centering
\begin{tabular}{ccccc}
\hline
\multicolumn{5}{c}{{\bf Flare forecasting using CNN obtained SHARPs features}}\\
\hline
\multicolumn{5}{c}{Number of observations}\\
\hline
\# Positives & \multicolumn{4}{c}{338}\\
\# Negatives & \multicolumn{4}{c}{6011}\\
\hline
& accuracy & recall(+) & recall(-) & TSS \\
\hline
True SHARPs      & 0.842 $\pm$ 0.030 & 0.856 $\pm$ 0.044 & 0.841 $\pm$ 0.033 & 0.697 $\pm$ 0.045 \\
CNN:HMI          & 0.812 $\pm$ 0.028 & 0.869 $\pm$ 0.056 & 0.809 $\pm$ 0.031 & 0.677 $\pm$ 0.046 \\
CNN:GONG         & 0.818 $\pm$ 0.031 & 0.801 $\pm$ 0.064 & 0.819 $\pm$ 0.035 & 0.621 $\pm$ 0.056 \\
True SHARPs \citep{bobraflareprediction} & 0.924 $\pm$ 0.007 & 0.832 $\pm$ 0.042 & 0.929 $\pm$ 0.008 & 0.761 $\pm$ 0.039\\
\hline
\end{tabular}
\caption{{\bf Flare-forecasting performance of the CNN-reconstructed vector-field features.} Flare-forecasting performance of a Support Vector Machine \citep{cortes1995support,hastie01statisticallearning} trained using CNN-estimated SHARPs features (Table~\ref{tab:dataNcorr}). The SVM is trained to forecast M- and X-class flares 24h in advance, similar to Bobra and Couvidat (2015) \citep{bobraflareprediction}. 1-$\sigma$ standard deviation is shown.}
\label{tab:flML}
\end{table*}
The SHARPs features considered (listed in Figure~\ref{fig:CNNModels} and Table~\ref{tab:dataNcorr}) are correlated among each other and are divided into four groups based on mutual Pearson correlations \citep{Dhuri2019}: (i) features that depend on the area of ARs, i.e., extensive features that include AR area, total unsigned flux, total unsigned vertical current, total unsigned current helicity, total free energy density and total Lorentz force, (ii) features that depend on the electric current in ARs, i.e., absolute net current helicity and sum of net current per polarity, (iii) features that depend only on the non-potential energy in ARs, i.e., mean free energy density and area with shear $>45\degree$, and finally, (iv) Schrijver R\_value \citep{Schrijver2007} viz. the sum of flux on the polarity inversion line. Overall, we develop four different CNNs (Figure~\ref{fig:CNNModels}) to estimate SHARPs features from these respective four groups. For each CNN, the output layer comprises $\rm K$ neurons to estimate $\rm K$ SHARPs features corresponding to each of the four groups.

Extensive features are strongly correlated with AR total unsigned flux, that depends only on the radial component of the magnetic field. The radial component is traditionally estimated from AR LOS magnetic field using a potential field approximation \citep{Leka2017}. Using a CNN, we directly estimate these extensive features without first requiring to estimate the radial magnetic field. The R\_value depends only on LOS magnetic field and can be directly calculated using GONG LOS magnetograms. However, to match HMI SHARPs R\_value, GONG LOS magnetic fields require a cross-calibration.  CNN models are expected to implicitly learn the cross-instrument calibration during training \citep{rs-713430} and estimated SHARPs values are also expected to be automatically cross-calibrated.

Unlike the extensive features, an accurate estimation of SHARPs depending on electric current and mean free energy requires explicit knowledge of the full vector-magnetic fields. Such features are important for understanding triggers of solar storms and are typically estimated assuming magnetic-field models, e.g., linear and non-linear force-free models \citep{Rgnier_2007}. Here, we provide a purely data-driven estimation of these features using a CNN. In order to assess the performance of the CNN, we use Linear Regression (LR) models as a baseline. We develop two separate LR models, one each for features that depend on electric current and free energy, respectively. As input, the LR models have extensive features and R\_Value. The first LR model produces absolute net current helicity and sum of net current per polarity as the output while the second produces mean free energy density and area with shear $> 45\degree$ as the output. Figure~\ref{fig:CNNModels} shows a schematic of the CNN models as well as the baseline models.

We use Pearson and Spearman correlations for measuring the performance of the CNN and baseline models. Pearson correlation measures a linear correlation between the true and estimated values of the vector-magnetic-field features. Spearman correlation is a rank correlation that captures the monotonic relationship between the true and estimated values in addition to the linear relationship measured by the Pearson correlation. Pearson and Spearman correlations for the CNN-estimated vector-magnetic-field features are listed in Tables~\ref{tab:dataNcorr}~and~\ref{tab:dataNcorrSp} respectively. For the baseline models, these correlations are listed in Table~\ref{tab:baseline}.

From Table~\ref{tab:dataNcorr}, the Pearson correlations of CNN-estimated vector-field features is higher for HMI than GONG and thus appear to be dependent on the spatial resolution of LOS magnetograms. For HMI, the CNN-estimated extensive features yield a Pearson correlation of $\sim 95\%$ for the validation and $\sim 90\%$ for the test data. For GONG data, the corresponding correlation is $\sim 90\%$. The Pearson correlations of CNN-estimated values of extensive features are not a perfect $\sim 100\%$, since the SHARPs calculation does not consider all pixels, rather, only taking into account those for which disambiguation of the azimuthal component of the magnetic field is reliable \citep{Bobra2014}. From Table~\ref{tab:dataNcorrSp}, the Spearman correlations for the extensive features are only slightly lower than the corresponding Pearson correlations implying that the ranking of the estimated features is generally consistent with the true ranking. 

The Pearson and Spearman correlation values for features that depend on the non-potential energy are significantly high $>90\%$ across the validation and test datasets. These correlation values are also $\sim 10\%-20\%$ higher compared to the linear regression baseline model (Table~\ref{tab:baseline}). These features, namely mean free energy density and area with shear $> 45\degree$, explicitly depend on the full vector-magnetic-field. 

For features that depend on electric current --- i.e., absolute net current helicity and sum of net current per polarity --- the CNN does not perform better than the baseline model. While the Pearson correlation for the validation are $20\%$ higher compared to the baseline of $70\%$, Spearman correlations are approximately equal (up to the error bars) at $60\%$. Also, the CNN fails to generalise to the test data with a low Pearson and Spearman correlation scores of $60\%$ each.

Figure~\ref{fig:corrNerror} shows scatter plot visualisations of the correlation between true and CNN-estimated SHARPs features for HMI and GONG from the ten validation sets. For HMI, the true and CNN-derived values mostly match relatively closely, except at only very small values ($< 200~G^2~-~m^{-1}$) of absolute net current helicity where the CNN estimates are significantly larger. For the HMI test data as well as GONG data, the CNN-estimation of absolute net current helicity for large values ($> 1000~G^2~-~m^{-1}$) is consistently on the lower side ($\sim 500~G^2~-~m^{-1}$). Figure~\ref{fig:mae} explicitly shows mean absolute errors in the CNN estimation as a function of the true values for HMI and GONG. Mean absolute errors in CNN-estimated values from GONG magnetograms show higher dependence on true values compared to HMI and increase significantly with increasing true values of the respective features, particularly for the validation data. For total unsigned flux, mean absolute errors of CNN-estimated features of both HMI and GONG are significantly higher for the extreme values $15-20~\times~10^{22}~{\rm Mx}$. For {the HMI test data and GONG data}, mean absolute errors are more than 12 times higher at large magnitudes ($> 1000~G^2~-~m^{-1}$) of absolute net current helicity compared to the HMI validation data. The average relative errors for GONG and HMI are comparable at $\approx80 \pm 10 \%$, $900 \pm 100\%$ and $25 \pm 2\%$ for total unsigned flux, absolute net current helicity and mean free energy density respectively. The high average relative errors imply that the CNN estimates are far off from true values, particularly for SHARPs features with low true values. SHARPs features from ARs which produce at least one major flare (M5 or greater) show a significant drop in average relative errors, at approximately $30 \pm 15 \%$, $300 \pm 80\%$ and $16 \pm 2\%$ respectively.

\subsection{Time evolution of the CNN-derived features on flaring active regions \label{sec:TE}}
For understanding AR magnetic-field dynamics and improving forecasting of solar storms, it is important that temporal variations of the CNN-estimated SHARPs is faithful to the true SHARPs. We measure trends in the time evolution of SHARPs features of an AR by fitting the observed and the CNN-estimated values with smooth spline curves and calculate numerical time derivatives. Table~\ref{tab:Trend_Measures} lists Pearson correlations between time derivatives of splines, fitted to the true and the CNN-estimated values of total unsigned flux, absolute net current helicity and mean free energy density. We find that the Pearson correlations are high, $\sim 80\%$, for HMI, with the exception of absolute net current helicity values from the test data.  For GONG, only the Pearson correlations for mean free energy density are high enough, $\sim 70\%$, to suggest that the corresponding trends are captured reasonably accurately in the CNN-estimated features. These discrepancies between trends of the CNN-estimated GONG and true values appear to be a consequence of the lower resolution of GONG magnetograms.

A comparison of the time evolution of true and CNN-estimated features obtained from HMI and GONG for individual ARs that produce at least one major flare (M5 or greater) is shown in Figure~\ref{fig:TE}. The true and CNN-estimated values of total unsigned flux, absolute net current helicity and mean free energy density are in agreement, particularly for HMI, capturing evolution of these features before and after flares. Disagreements between the true and CNN-estimated features occur only at the extreme values of these features.  E.g., for X9.3 flare in NOAA 12673 in September 2017, which was the largest flare in cycle 24, the CNN accurately estimates the rise of the total unsigned flux and also mean free energy density prior to the flare. The absolute net current helicity rises to unusually high values prior to the X9.3 flare ($30\%$ more than the maximum values encountered in the training data) and therefore the corresponding CNN estimates are inaccurate. More examples of comparisons of the time evolution between true and CNN-estimated features from ARs that produce at least one M5 or greater flare are included in the Appendix~Figure~\ref{fig:allTE}.

The CNN estimation of SHARPs features on flaring ARs is thus useful for understanding AR magnetic-field evolution leading to particularly violent solar storms in the past. The Halloween storms of October 2003 produced extreme flares from AR NOAA 10486 of magnitudes X17.0, X10.0 and the largest recorded flare X28.0 \citep{Pulkkinen2005}. The magnetic-field evolution leading to these extreme flares was characterised by rotation of a major positive polarity of the delta sunspot as shown in the top panel of Figure~\ref{fig:timeEvoleHalloween} \citep{Zhang2008}. Without the knowledge of vector-magnetic-fields, free energy and current helicity during these storms are previously modelled based on the magnetic virial theorem \citep{Metcalf_2005,Rgnier_2007}, linear/non-linear force-free field extrapolation \citep{Rgnier_2007}, and a Minimum Current Corona model \citep{Kazachenko_2010}. We obtain a model-free and purely data-driven CNN-estimates of total unsigned flux, absolute net current helicity and mean free energy density during these storms using LOS magnetograms. However, the HMI observations are not available for this period. We therefore use the CNN trained with HMI magnetograms to process LOS observations from MDI during the Halloween storms to estimate time evolution of total unsigned flux, absolute net current helicity and mean free energy density. Flare X28.0 is excluded as it occured outside $45\degree$ of the central meridian. In particular, the CNN-estimated absolute net current helicity of NOAA 10486 rises continuously by $25\%$ between X1.2 flare and X17.0 flare corresponding to the observed sunspot rotation.  A similar gradual rise of a modelled helicity flux by $50\%$ between X1.2 and X17.0 flare has been reported  \citep{Kazachenko_2010}, caused primarily by helicity injection from the rotation of the sunspot. The CNN estimates show that the absolute net current helicity stays high leading to the X10.0 flare and falls thereafter. The CNN-estimated mean free energy density also rises leading to the X17.0 and X10.0 flares. 
Note that these CNN-estimated values from MDI magnetograms are not expected to be corrected for the instrument cross-calibration between the MDI and HMI since the CNN is trained with only HMI magnetograms. Table~\ref{tab:dataNcorrMDI} in the Appendix lists the Pearson and Spearman correlations between the true values and the CNN-estimated values using MDI line-of-sight magnetograms, during the overlap period of MDI and HMI. These correlation values are significantly lower compared to those estimated from HMI magnetograms (Tables~\ref{tab:dataNcorr} and \ref{tab:dataNcorrSp}). Therefore, a rigorous estimation first requires standardisation of MDI and HMI magnetograms (e.g., with other approaches such as super-resolution \citep{rs-713430}). We also used the GONG magnetograms to estimate the vector-field features during the storms using the CNN trained with GONG (see Appendix Figure~\ref{fig:allHallow}). The values of the vector-field features estimated using GONG magnetograms are in the extreme range, as expected during the storms.  However, the sensitivity of these estimated values to pre- and post-flare magnetic-field variations is lower compared to the features estimated from MDI.

\subsection{Flare forecasting using CNN-derived features}
\begin{figure*}[t]
\centering
\subfloat{\includegraphics[width= 0.33\textwidth,trim={0.0cm 0.0cm 0.0cm 0.0cm},clip]{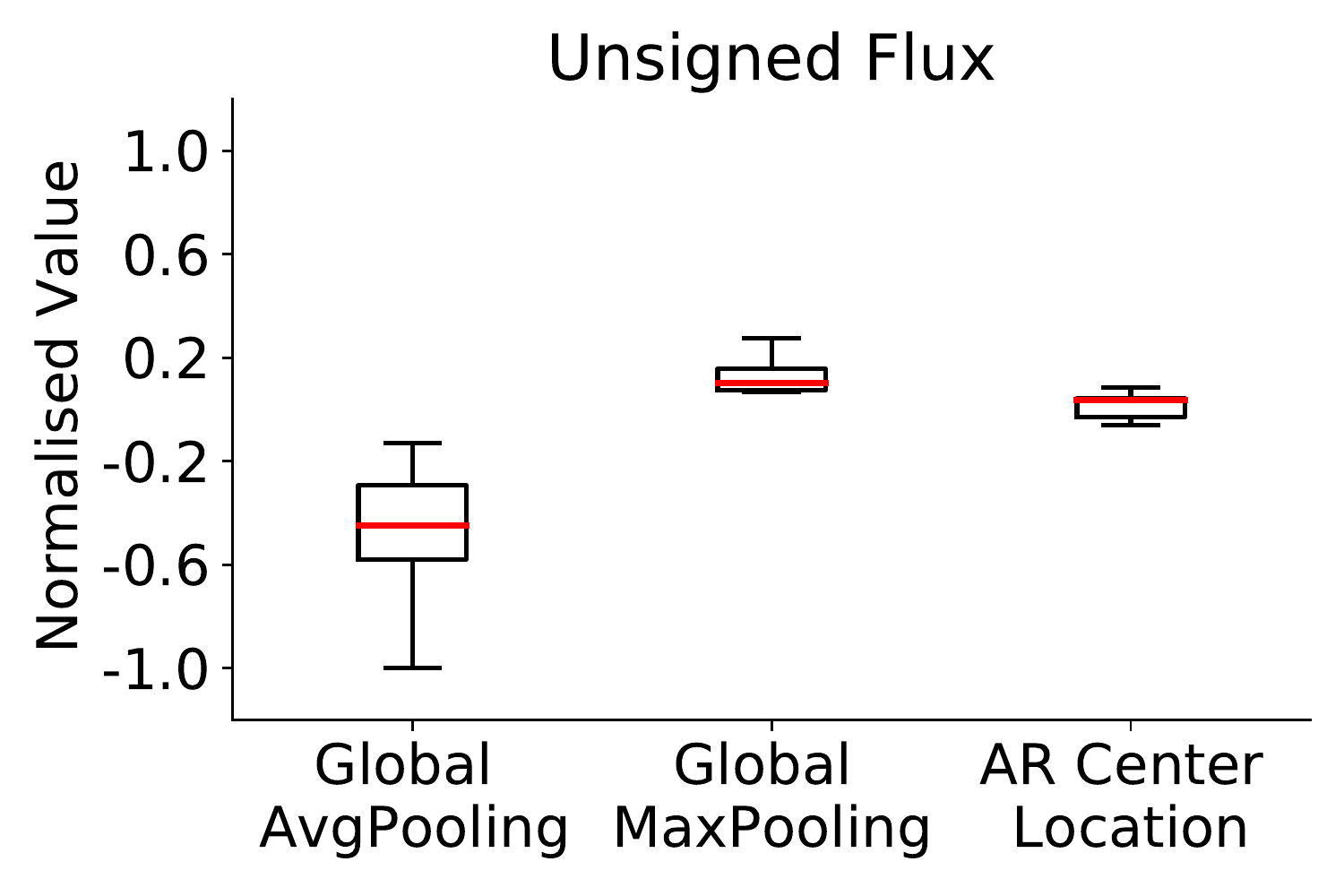}}
\subfloat{\includegraphics[width= 0.33\textwidth,trim={0.0cm 0.0cm 0.0cm 0.0cm},clip]{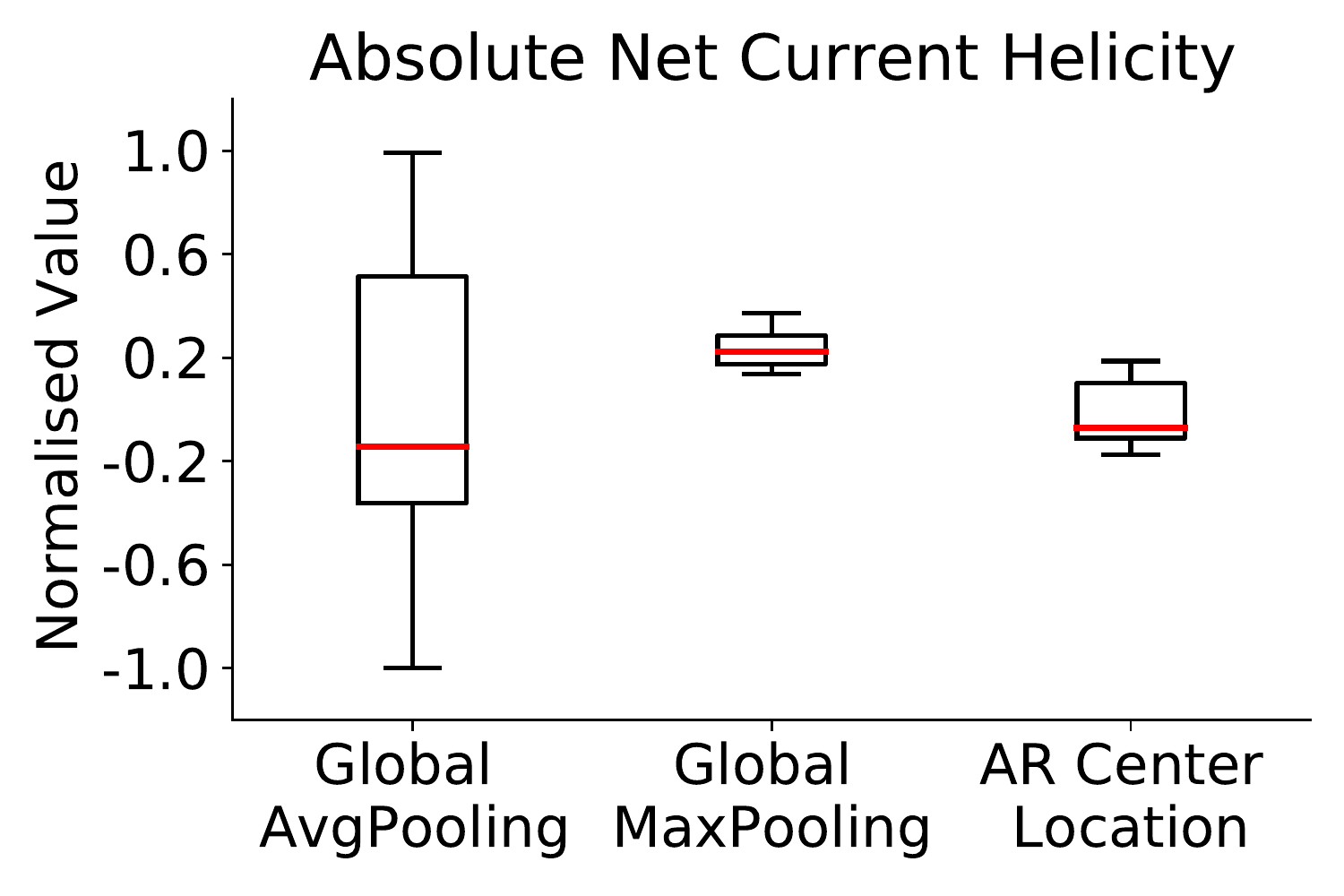}}
\subfloat{\includegraphics[width= 0.33\textwidth,trim={0.0cm 0.0cm 0.0cm 0.0cm},clip]{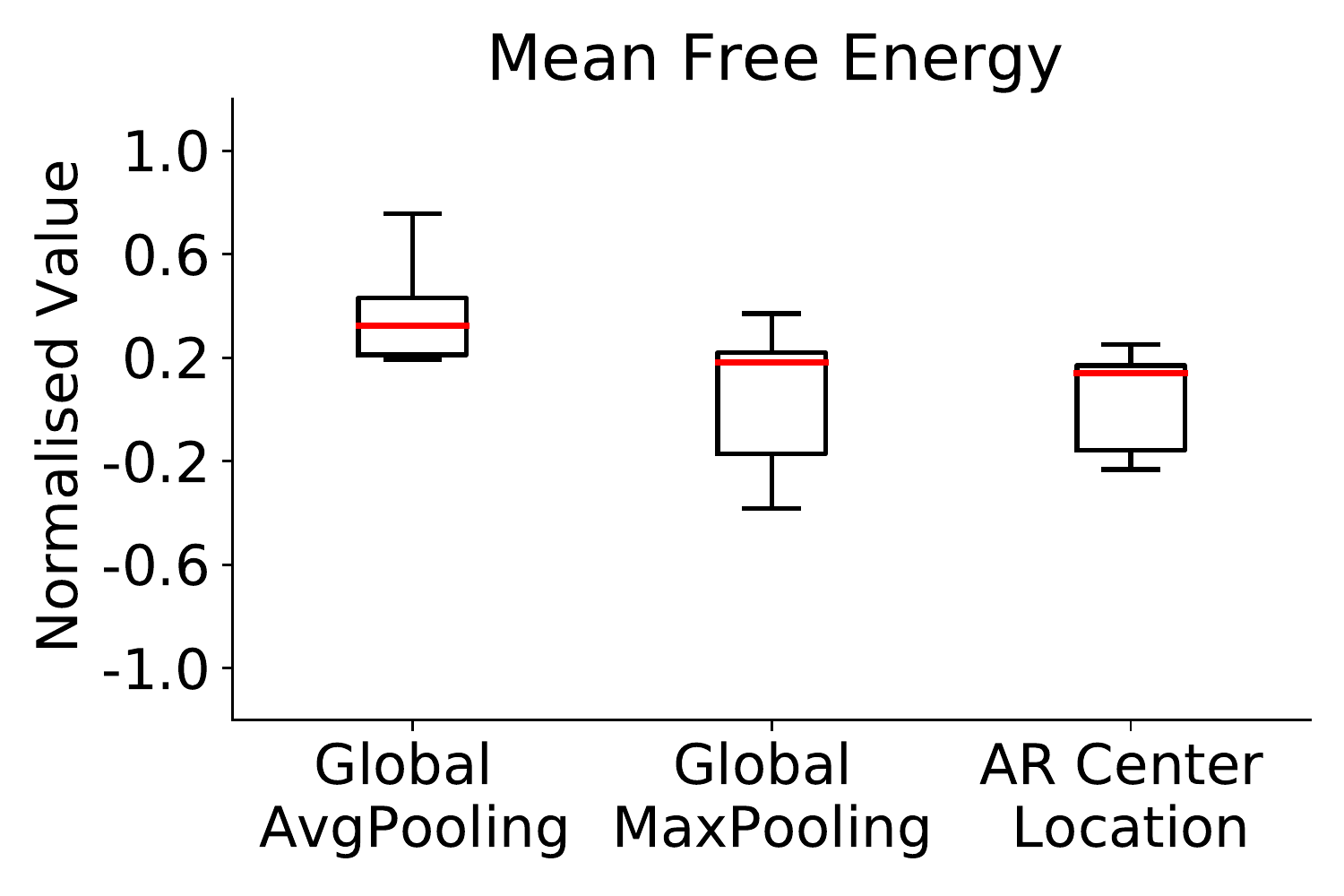}}\\
\caption{{\bf Interpretation of the trained CNN model.} Boxplots showing weights in the penultimate layer of the CNN trained to reconstruct total unsigned flux, absolute net current helicity and mean free energy density. Top five weights from the components of the penultimate layer, i.e.. global average pooling, global max-pooling and fully connected layer processing AR location (see Figure~\ref{fig:CNN}), are shown for the ten validation models. The red line indicates median of the weight populations.}
\label{fig:wTSummary}
\end{figure*}
\begin{figure*}
\centering
\includegraphics[width= 0.68\textwidth,trim={1.75cm 1.2cm 2.0cm 0.7cm},clip]{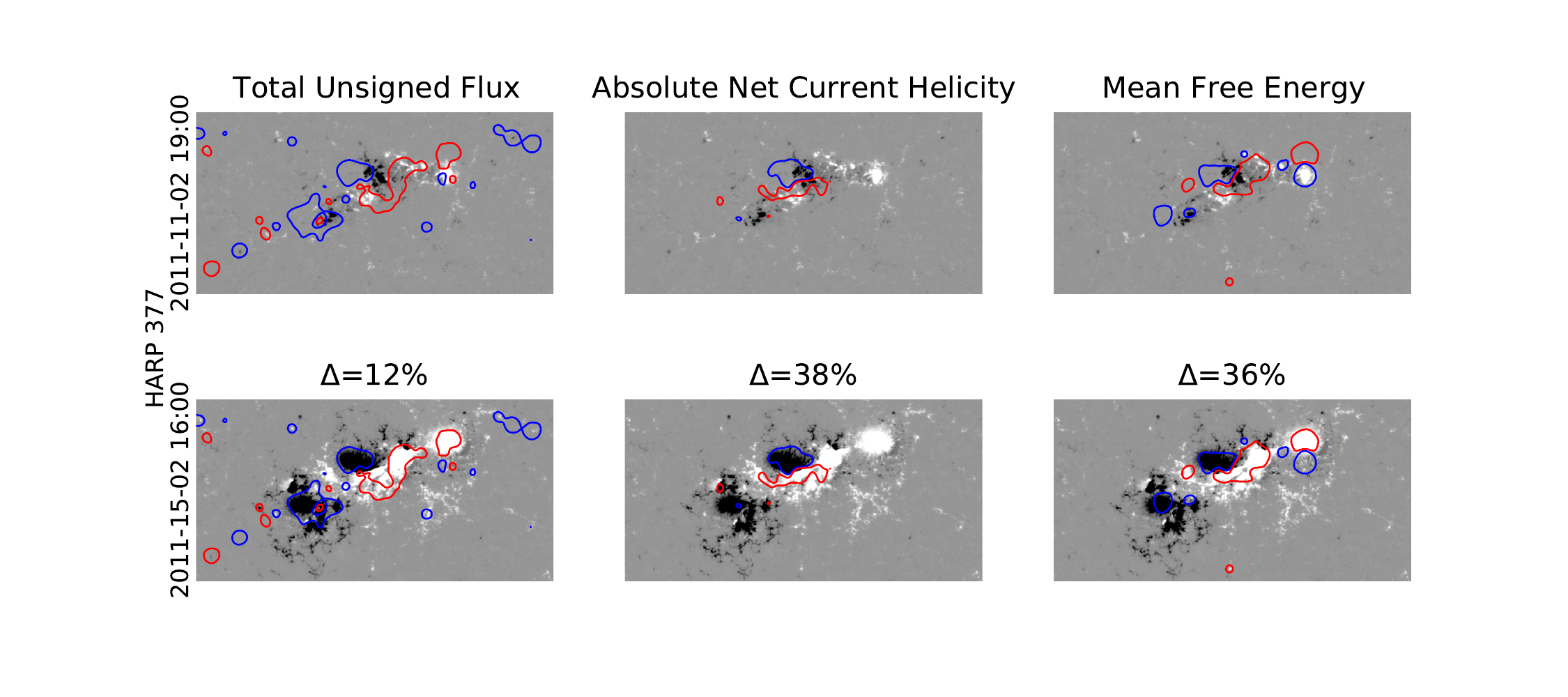}\\
\vspace{\baselineskip}
\includegraphics[width= 0.68\textwidth,trim={1.75cm 1.2cm 2.0cm 0.7cm},clip]{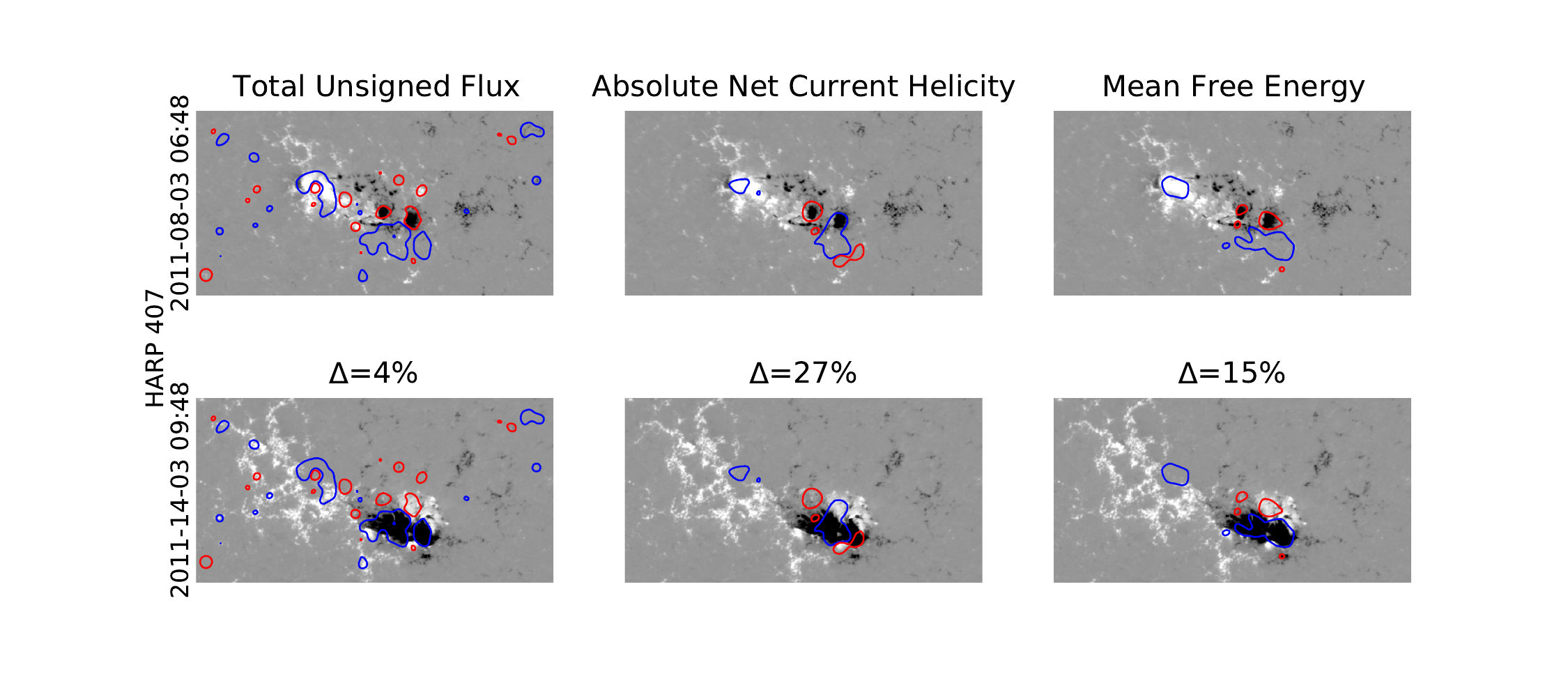}\\
\vspace{\baselineskip}
\includegraphics[width= 0.68\textwidth,trim={1.75cm 1.2cm 2.0cm 0.7cm},clip]{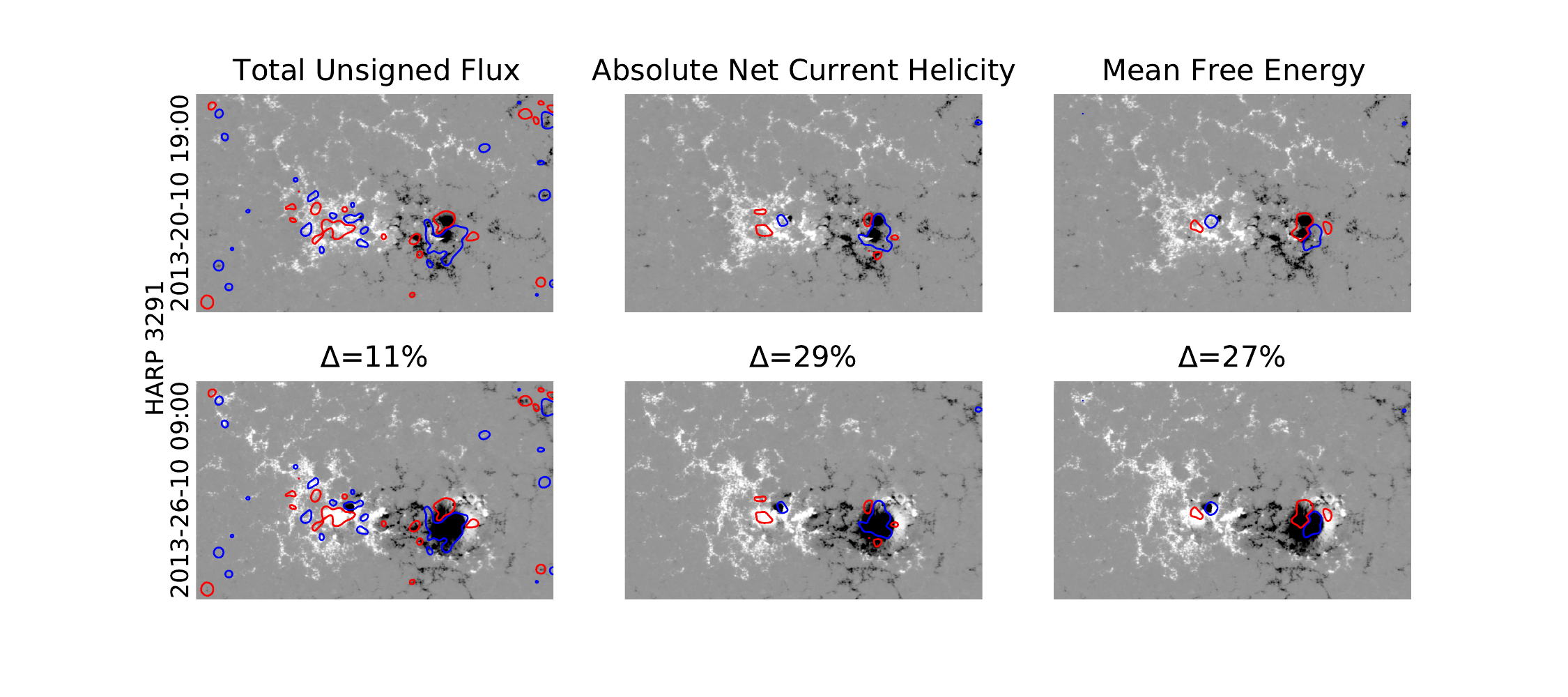}\\
\vspace{\baselineskip}
\includegraphics[width= 0.68\textwidth,trim={1.75cm 1.2cm 2.0cm 0.7cm},clip]{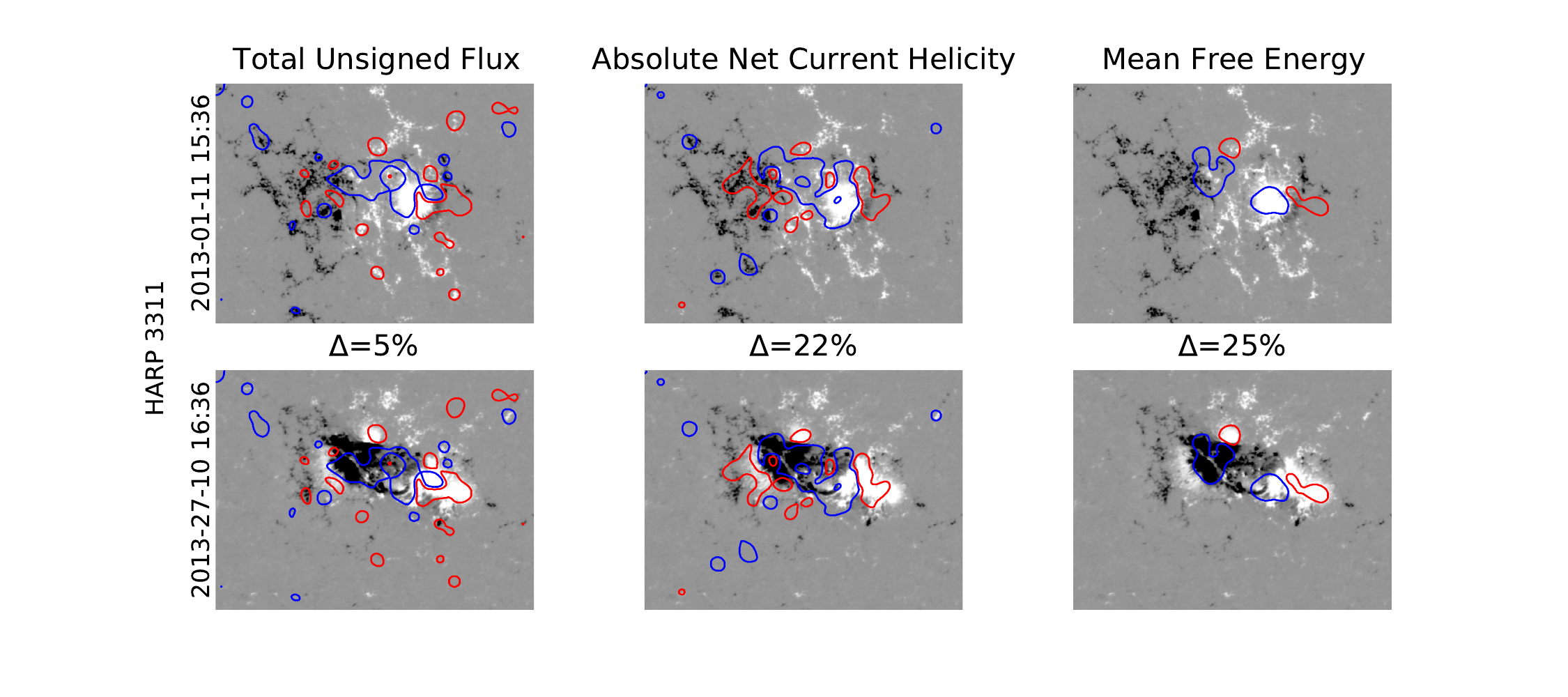}
\caption{{\bf Integrated gradient (IG) attribution maps}. The contour plots for the IG attribution maps highlight the regions from magnetograms with the highest attributions for the CNN estimation of SHARPs features --- total unsigned flux, absolute net current helicity, and mean free energy density. The red/blue contours correspond to the regions with the net positive/negative attributions for the magnetograms in the bottom rows relative to the reference magnetograms in the top rows. The $\Delta$ values are the percentage change in the normalised values of the respective features between the present and the reference observations. The colorbar for the magnetograms is saturated with $\rm{\pm~500~G}$. The IG attribution maps are smoothed with a Gaussian filter of standard deviation 10 pixels before obtaining the contours.} 
\label{fig:IG}
\end{figure*}
The SHARPs features have been extensively used for building flare-forecasting models using ML \citep{bobraflareprediction,Bobra2016,Nishizuka2017,Dhuri2019,Chen2019,Ahmadzadeh_2021}. In order to assess the utility of CNN-estimated SHARPs for flare-forecasting tasks, we compare their flare-forecasting performance to the true SHARPs. We set up the problem of forecasting M-/X-class flares with 24h warning similar to \citet{bobraflareprediction}.  We use two approaches for the comparison. First, we build Linear Discriminant Analysis (LDA) classification models using one SHARPs feature at a time. This allows for direct comparing of the true and the CNN-estimated values of each SHARPs feature for flare forecasting. Second, we use all SHARPs features together to train a support vector machine (SVM) for flare forecasting. We measure the flare-forecasting performance using accuracy, recall and the True Skill Statistics (TSS) score \citep{PEIRCE453}. However, only the latter two are robust to the class-imbalance prevalent in the flare-forecasting problem \citep{bobraflareprediction,Ahmadzadeh_2021}, and therefore reliable for comparison. Our definitions of positive and negative classes are identical to the operational approach described in \citet{bobraflareprediction}. In addition, we use the 10-times repeated-holdout validation described in Section~\ref{sec:Methods}. Unlike \citet{bobraflareprediction}, we explicitly ensure that the samples from a given AR are not mixed in training and validation sets \citep{Ahmadzadeh_2021}. Also, as mentioned in Section~\ref{sec:Data}, we only consider ARs with maximum area $>25~{\rm Mm^2}$. Both the LDA and SVM are implemented using the scikit-learn library in Python.

Table~\ref{tab:LDA} lists performance metrics for the classification of M-/X-class flares using the LDA of one SHARPs feature at a time. The accuracy, recall and TSS values obtained using each of the CNN-estimated features from HMI and GONG magnetograms are consistent with those of the true SHARPs features up to the validation error bars. We note that Schrijver's R\_value \citep{Schrijver2007} gives the highest TSS values for flare forecasting using individual features.

Table~\ref{tab:flML} lists the performance metrics for the SVM classification of M-/X-class flares using all SHARPs features together. TSS ($\sim 68\%$) and recall ($\sim 86\%$) values obtained using an SVM trained with the CNN-estimated features from HMI are consistent with those obtained using the true SHARPs. TSS ($\sim 62\%$) and recall ($\sim 80\%$) values from an SVM trained with the CNN-estimated features from GONG are slightly lower. For a comparison, we list TSS ($\sim 76\%$) and recall ($\sim 83\%$) from \citet{bobraflareprediction} that are higher. The systematically lower TSS of the SVM in forecasting flares when using true SHARPs values here as compared with \citet{bobraflareprediction} is due to exclusion of observations from ARs with maximum area $<25~{\rm Mm^2}$ (all nonflaring) and the explicit restriction that samples from an AR are part of either training or validation sets. Largely consistent performance metrics for flare forecasting with the CNN-estimated SHARPs imply that high relative errors notwithstanding, the CNN-estimated features can be useful for building space-weather forecasting tools. This is a consequence of (true) SHARPs feature values varying over several orders of magnitudes and thus being significantly different for flaring and nonflaring ARs for forecasting of flares \citep{Dhuri2019}. Accuracy of the CNN-estimated SHARPs features may be improved by significantly increasing the resolution of LOS magnetograms from, e.g., GONG, using techniques such as super-resolution \citep{rs-713430}. Our method is thus suitable for reconstructing vector-field features from historical LOS magnetograms, ultimately useful for reliable space-weather forecasting.

\subsection{Interpreting the CNN}
CNNs and, in general, deep learning are extremely efficient at identifying  correlations in the data. In this case, the CNN builds a useful model of AR vector magnetic fields from the observed LOS magnetograms. In particular, the CNN estimated SHARPs features may be reliably used to study energy build up and time evolution of magnetic fields in flaring ARs. Yet it is very challenging to open up the trained network and understand the CNN to uncover the information absorbed. Nevertheless, weights learned by the CNN can shed some light on its working. There are also attribution methods to quantify the contribution of different parts of the input image to the CNN's output. Here, we analyse the weights of the CNN as well as obtain attribution maps for input magnetograms to interpret the trained CNN.

The CNN architecture (Figure~\ref{fig:CNN}) comprises a fully convolutional network for processing LOS magnetograms and a fully connected layer of neurons for processing information about the location of ARs on the solar disk. The penultimate concatenation layer comprises a global-average-pooling layer, a global-max-pooling layer that process the LOS magnetograms and a fully-connected layer that processes the location of the ARs. The global-average-pooling neurons are sensitive to the entire spatial extent of LOS magnetograms, while the global-max-pooling neurons are sensitive to spatially local patterns. The fully-connected neurons are sensitive to AR coordinates on the disk. Figure~\ref{fig:wTSummary} illustrates the distribution of the top weights of each of the three components in the penultimate layers as their contribution to the output of the CNN that estimates total unsigned flux, absolute net current helicity and mean free energy density. Neurons associated with global-average pooling contribute dominantly to the total unsigned flux and mean free energy, implying that their estimation depends on the consideration of the entire LOS magnetograms. For absolute net current helicity, key contributors are neurons from the global max-pooling layer and its estimation is sensitive to spatially local patterns from LOS magnetograms. Without the global-max-pooling layer, absolute net current helicity and related CNN-estimated SHARPs features show $\sim 30\%$ less Pearson correlation with the true values. Weights from neurons related to AR location on the solar disk are $\sim$ 0, and thus, the CNN estimation does not strongly depend on the AR location. Indeed, the CNN may be trained equally well without the additional input of the AR location. This may be a consequence of considering AR patches only within $\pm 45\degree$  where the projection effects are not significant.

While there are many attribution methods, gradient-based methods such as saliency maps \citep{Simonyan2013}, grad-CAMs \citep{Selvaraju2017}, integrated gradients (IG) \citep{pmlr-v70-sundararajan17a} etc., are favoured over perturbation-based methods such as occlusion masks \citep{Zeiler2014} because of computational efficiency and higher resolution attribution maps. IG attribution maps are of the same resolution as the input magnetograms and are thus superior to grad-CAMs obtained from the CNN feature maps. Also, unlike saliency maps, IG attribution maps are calculated using a reference input image that facilitates assigning a cause for the attribution e.g. by comparing the magnetic-field evolution \citep{Sun_2022}. Thus, here we use IG attribution maps to identify pixels, and hence the magnetic-field features in the input, that are important for the CNN output. The IG attribution map for a given input image is calculated by integrating gradients in the CNN output along the path from a reference image. Formally,

\begin{equation}
L^{f} \left(x,x_{0}\right) = \left(x-x_{0}\right) \times \int_{\alpha=0}^{1} \frac{\partial Y^{f}\left(x_{0} + \alpha \times \left(x-x_{0}\right)\right)}{\partial x}d\alpha,
 \end{equation}
where $x_{0}$ is the reference image, $Y^{f}$ is the CNN output for SHARPs vector-field feature $f$.

Figure~\ref{fig:IG} shows contour plots of typical IG attribution maps for a few example magnetograms from flaring ARs ({\it bottom rows}). The red/blue contours include regions of net positive/negative contribution towards the CNN output. The IG attribution maps for the three SHARPs features --- total unsigned flux, absolute net current helicity, and mean free energy density --- are shown separately along with the reference magnetograms ({\it top rows}) used. In general, increasing/decreasing positive polarity flux corresponds to net positive/negative attribution. For total unsigned flux, almost all magnetic-field regions, even relatively smaller regions with weaker magnetic fields, constitute a positive/negative attribution. In contrast, for absolute net current helicity and mean free energy density, only relatively larger and stronger magnetic-field regions constitute a positive/negative attribution. For mean free energy, positive/negative attribution regions typically correspond to the uniformly increasing/decreasing positive flux. In the case of absolute net current helicity, attributions correspond to regions with ''mixed" magnetic fields of the positive-negative polarities closely located. The appearance of a spurious magnetic-field polarity inversion line (PIL) is a known artifact in the line-of-sight magnetograms whenever the magnetic-field inclination relative to the line-of-sight exceeds ${\textrm 90\degree}$ \citep{Leka2017}. We find that in many cases (e.g. HARPs 407, 3291, and 3311) when the PIL artifact exists for magnetic fields within penumbrae, it wrongly constitutes an important attribution. The misattribution results from the failure of the CNN to learn the PIL artifact \citep{Sun_2022} and as a consequence, limits the accuracy of the reconstructed the vector-field-features.

\section{Discussion}
We have thus developed a CNN model for quantifying vector-field properties --- extensive features such as total unsigned flux as well as properties depending explicitly on transverse magnetic-field component such as free-energy density and current helicity --- using LOS magnetograms taken from space-based HMI and ground-based GONG instruments. The CNN-estimated features strongly correlate ($> 90\%$) with their true measurements from HMI SHARPs, particularly for high-resolution LOS magnetograms from HMI. Time-evolution of the CNN-estimated features reliably mimic true AR magnetic-field evolution, particularly for ARs producing major flares (M5 or greater). Prior to HMI, vector-magnetic-field observations available from instruments such as Imaging Vector Magnetograph and Hinode/Spectro Polarimeter \citep{Kosugi2007} have limited spatial and temporal converge. In contrast, near-continuous observations of LOS magnetograms are available since the 1970s from missions such as the Kitt Peak telescope (KP), MDI and GONG. LOS magnetograms from these instruments vary in their spatial resolution that are lower than HMI resolution. Nonetheless, these instruments' observation periods overlap with HMI (KP:2010-present, MDI:2010-2011, GONG:2010-present) and the attendant observations may be used to train or fine tune the CNN model to estimate SHARPs vector-field features. We explicitly show that the flare-forecasting performance of the CNN-estimated features is comparable to the true SHARPs. Therefore, vector-fields estimated from past LOS observations of nearly five decades using CNN can provide approximately four times more solar storms' data than currently available, useful for building robust statistical models for space-weather forecasting using ML. A larger sample size of solar storms also facilitates building ML algorithms based on time series of AR observations which may significantly improve forecasting performance \citep{Dhuri2019}. The CNN estimated vector-fields also provide a new perspective to understand  and quantify magnetic-field dynamics during the past extreme events such as 2003 Halloween storms as demonstrated here.

Our CNN estimates are reliable for studies of solar storms, yet there is also a significant scope of improvement. Our estimates of vector-field features using HMI magnetograms are consistently more accurate compared to those estimated using lower resolution GONG magnetograms. Using LOS magnetograms from GONG and other instruments that are explicitly cross-calibrated with HMI LOS magnetograms may significantly improve accuracy of the corresponding vector-field instruments. Also, Deep-learning-based techniques for improving the resolution of magnetograms, namely super-resolution, are being successfully developed \citep{Rahman_2020,rs-713430}. Using super-resolved LOS magnetograms as input to the CNN promises to yield more accurate CNN estimates of the vector-field features. Our estimates are also based only on the training data from the rising phase of cycle 24.  Using new data available from HMI and also from newer instruments, a robust CNN regression is achievable. Extending our method, reasonable data-driven estimates of even the full photospheric vector-magnetic-field from only LOS magnetograms may be feasible, which opens up a new approach in studying and modelling AR magnetic-fields using ML.

\section*{Acknowledgments}
S.M.H acknowledges funding from Department of Atomic Energy grant RTI4002 and the Max-Planck Partner Group programme. D.B.D and S.M.H. acknowledge discussions with Mark C. M. Cheung and Marc DeRosa. The authors would also like to thank the anonymous reviewer and the scientific editor  Manolis K. Georgoulis for their comments and suggestions that helped improve clarity of the manuscript. The authors declare that they have no competing interests. D.B.D. and S.M.H. designed the research. D.B.D., S.B. and S.K.M. analysed data. D.B.D. and S.M.H. interpreted the results. D.B.D. wrote the manuscript with contributions from S.M.H. HMI LOS and vector magnetograms, MDI LOS magnetograms and SHARPs data are publicly accessible on the JSOC data server at \href{http://jsoc.stanford.edu/}{http://jsoc.stanford.edu/}, courtesey the HMI and MDI science teams. The GONG LOS magnetograms are publicly available at \href{https://gong.nso.edu/}{https://gong.nso.edu/} and were acquired by GONG instruments operated by NISP/NSO/AURA/NSF with contribution from NOAA. 

\bibliographystyle{aasjournal}
\bibliographystyle{aasjournal}

\begin{thebibliography}{}
\expandafter\ifx\csname natexlab\endcsname\relax\def\natexlab#1{#1}\fi
\providecommand{\url}[1]{\href{#1}{#1}}
\providecommand{\dodoi}[1]{doi:~\href{http://doi.org/#1}{\nolinkurl{#1}}}
\providecommand{\doeprint}[1]{\href{http://ascl.net/#1}{\nolinkurl{http://ascl.net/#1}}}
\providecommand{\doarXiv}[1]{\href{https://arxiv.org/abs/#1}{\nolinkurl{https://arxiv.org/abs/#1}}}

\bibitem[{Ahmadzadeh {et~al.}(2021)Ahmadzadeh, Aydin, Georgoulis, Kempton,
  Mahajan, \& Angryk}]{Ahmadzadeh_2021}
Ahmadzadeh, A., Aydin, B., Georgoulis, M.~K., {et~al.} 2021, The Astrophysical
  Journal Supplement Series, 254, 23, \dodoi{10.3847/1538-4365/abec88}

\bibitem[{Bhattacharjee {et~al.}(2020)Bhattacharjee, Alshehhi, Dhuri, \&
  Hanasoge}]{Bhattacharjee2020}
Bhattacharjee, S., Alshehhi, R., Dhuri, D.~B., \& Hanasoge, S.~M. 2020, The
  Astrophysical Journal, 898, 98, \dodoi{10.3847/1538-4357/ab9c29}

\bibitem[{Bobra \& Couvidat(2015)}]{bobraflareprediction}
Bobra, M.~G., \& Couvidat, S. 2015, The Astrophysical Journal, 798, 135.
\newblock \url{http://stacks.iop.org/0004-637X/798/i=2/a=135}

\bibitem[{{Bobra} \& {Ilonidis}(2016)}]{Bobra2016}
{Bobra}, M.~G., \& {Ilonidis}, S. 2016, The Astrophysical Journal, 821, 127,
  \dodoi{10.3847/0004-637X/821/2/127}

\bibitem[{Bobra {et~al.}(2014)Bobra, Sun, Hoeksema, Turmon, Liu, Hayashi,
  Barnes, \& Leka}]{Bobra2014}
Bobra, M.~G., Sun, X., Hoeksema, J.~T., {et~al.} 2014, Solar Physics, 289,
  3549, \dodoi{10.1007/s11207-014-0529-3}

\bibitem[{Bobra {et~al.}(2021)Bobra, Wright, Sun, \& Turmon}]{Bobra_2021}
Bobra, M.~G., Wright, P.~J., Sun, X., \& Turmon, M.~J. 2021, The Astrophysical
  Journal Supplement Series, 256, 26, \dodoi{10.3847/1538-4365/ac1f1d}

\bibitem[{Boteler(2019)}]{Boteler2019}
Boteler, D.~H. 2019, Space Weather, 17, 1427, \dodoi{10.1029/2019SW002278}

\bibitem[{Bottou(1991)}]{Bottou91stochasticgradient}
Bottou, L. 1991, in Proceedings of Neuro-Nimes, 687--706

\bibitem[{Chen {et~al.}(2019)Chen, Manchester, Hero, Toth, DuFumier, Zhou,
  Wang, Zhu, Sun, \& Gombosi}]{Chen2019}
Chen, Y., Manchester, W.~B., Hero, A.~O., {et~al.} 2019, Space Weather, 17,
  1404, \dodoi{10.1029/2019SW002214}

\bibitem[{Cheung \& Isobe(2014)}]{Cheung2014}
Cheung, M. C.~M., \& Isobe, H. 2014, Living Reviews in Solar Physics, 11, 3,
  \dodoi{10.12942/lrsp-2014-3}

\bibitem[{Cortes \& Vapnik(1995)}]{cortes1995support}
Cortes, C., \& Vapnik, V. 1995, Machine learning, 20, 273

\bibitem[{Crown(2012)}]{crown2012validation}
Crown, M.~D. 2012, Space Weather, 10, 1, \dodoi{10.1029/2011SW000760}

\bibitem[{Dhuri {et~al.}(2019)Dhuri, Hanasoge, \& Cheung}]{Dhuri2019}
Dhuri, D.~B., Hanasoge, S.~M., \& Cheung, M. C.~M. 2019, Proceedings of the
  National Academy of Sciences, 116, 11141, \dodoi{10.1073/pnas.1820244116}

\bibitem[{Eastwood {et~al.}(2017)Eastwood, Biffis, Hapgood, Green, Bisi,
  Bentley, Wicks, McKinnell, Gibbs, \& Burnett}]{Eastwood2017}
Eastwood, J.~P., Biffis, E., Hapgood, M.~A., {et~al.} 2017, Risk Analysis, 37,
  206, \dodoi{10.1111/risa.12765}

\bibitem[{Goodfellow {et~al.}(2016)Goodfellow, Bengio, \&
  Courville}]{Goodfellow2016}
Goodfellow, I., Bengio, Y., \& Courville, A. 2016, Deep Learning (The MIT
  Press)

\bibitem[{Han \& Moraga(1995)}]{10.1007/3-540-59497-3_175}
Han, J., \& Moraga, C. 1995, in From Natural to Artificial Neural Computation,
  ed. J.~Mira \& F.~Sandoval (Berlin, Heidelberg: Springer Berlin Heidelberg),
  195--201

\bibitem[{Hastie {et~al.}(2001)Hastie, Tibshirani, \&
  Friedman}]{hastie01statisticallearning}
Hastie, T., Tibshirani, R., \& Friedman, J. 2001, The Elements of Statistical
  Learning, Springer Series in Statistics (New York, NY, USA: Springer New York
  Inc.)

\bibitem[{Hoeksema {et~al.}(2014)Hoeksema, Liu, Hayashi, Sun, Schou, Couvidat,
  Norton, Bobra, Centeno, Leka, Barnes, \& Turmon}]{Hoeksema2014}
Hoeksema, J.~T., Liu, Y., Hayashi, K., {et~al.} 2014, Solar Physics, 289, 3483,
  \dodoi{10.1007/s11207-014-0516-8}

\bibitem[{Kazachenko {et~al.}(2010)Kazachenko, Canfield, Longcope, \&
  Qiu}]{Kazachenko_2010}
Kazachenko, M.~D., Canfield, R.~C., Longcope, D.~W., \& Qiu, J. 2010, The
  Astrophysical Journal, 722, 1539, \dodoi{10.1088/0004-637x/722/2/1539}

\bibitem[{Kosugi {et~al.}(2007)Kosugi, Matsuzaki, Sakao, Shimizu, Sone,
  Tachikawa, Hashimoto, Minesugi, Ohnishi, Yamada, Tsuneta, Hara, Ichimoto,
  Suematsu, Shimojo, Watanabe, Shimada, Davis, Hill, Owens, Title, Culhane,
  Harra, Doschek, \& Golub}]{Kosugi2007}
Kosugi, T., Matsuzaki, K., Sakao, T., {et~al.} 2007, Solar Physics, 243, 3,
  \dodoi{10.1007/s11207-007-9014-6}

\bibitem[{LeCun {et~al.}(2015)LeCun, Bengio, \& Hinton}]{LeCun2015}
LeCun, Y., Bengio, Y., \& Hinton, G. 2015, Nature, 521, 436 EP .
\newblock \url{https://doi.org/10.1038/nature14539}

\bibitem[{Leka \& Barnes(2007)}]{Leka2008}
Leka, K.~D., \& Barnes, G. 2007, The Astrophysical Journal, 656, 1173.
\newblock \url{http://stacks.iop.org/0004-637X/656/i=2/a=1173}

\bibitem[{Leka {et~al.}(2017)Leka, Barnes, \& Wagner}]{Leka2017}
Leka, K.~D., Barnes, G., \& Wagner, E.~L. 2017, Solar Physics, 292, 36,
  \dodoi{10.1007/s11207-017-1057-8}

\bibitem[{Livingston {et~al.}(1976)Livingston, Harvey, Pierce, Schrage,
  Gillespie, Simmons, \& Slaughter}]{Livingston:76}
Livingston, W.~C., Harvey, J., Pierce, A.~K., {et~al.} 1976, Appl. Opt., 15,
  33, \dodoi{10.1364/AO.15.000033}

\bibitem[{McIntosh(1990)}]{McIntosh1990}
McIntosh, P.~S. 1990, Solar Physics, 125, 251, \dodoi{10.1007/BF00158405}

\bibitem[{Metcalf {et~al.}(2005)Metcalf, Leka, \& Mickey}]{Metcalf_2005}
Metcalf, T.~R., Leka, K.~D., \& Mickey, D.~L. 2005, The Astrophysical Journal,
  623, L53, \dodoi{10.1086/429961}

\bibitem[{Munoz-Jaramillo {et~al.}(2022)Munoz-Jaramillo, Jungbluth, Gitiaux,
  Wright, Shneider, Maloney, Baydin, Gal, Deudon, \& Kalaitzis}]{rs-713430}
Munoz-Jaramillo, A., Jungbluth, A., Gitiaux, X., {et~al.} 2022, Nature
  Portfolio, \dodoi{10.21203/rs.3.rs-713430/v1}

\bibitem[{Nishizuka {et~al.}(2017)Nishizuka, Sugiura, Kubo, Den, Watari, \&
  Ishii}]{Nishizuka2017}
Nishizuka, N., Sugiura, K., Kubo, Y., {et~al.} 2017, The Astrophysical Journal,
  835, 156.
\newblock \url{http://stacks.iop.org/0004-637X/835/i=2/a=156}

\bibitem[{Peirce(1884)}]{PEIRCE453}
Peirce, C.~S. 1884, Science, ns-4, 453, \dodoi{10.1126/science.ns-4.93.453-a}

\bibitem[{Pesnell {et~al.}(2012)Pesnell, Thompson, \&
  Chamberlin}]{Pesnell-etall2012}
Pesnell, W.~D., Thompson, B.~J., \& Chamberlin, P.~C. 2012, Solar Physics, 275,
  3, \dodoi{10.1007/s11207-011-9841-3}

\bibitem[{Pulkkinen {et~al.}(2005)Pulkkinen, Lindahl, Viljanen, \&
  Pirjola}]{Pulkkinen2005}
Pulkkinen, A., Lindahl, S., Viljanen, A., \& Pirjola, R. 2005, Space Weather,
  3, 1, \dodoi{10.1029/2004SW000123}

\bibitem[{Rahman {et~al.}(2020)Rahman, Moon, Park, Siddique, Cho, \&
  Lim}]{Rahman_2020}
Rahman, S., Moon, Y.-J., Park, E., {et~al.} 2020, The Astrophysical Journal,
  897, L32, \dodoi{10.3847/2041-8213/ab9d79}

\bibitem[{R{\'{e}}gnier \& Priest(2007)}]{Rgnier_2007}
R{\'{e}}gnier, S., \& Priest, E.~R. 2007, The Astrophysical Journal, 669, L53,
  \dodoi{10.1086/523269}

\bibitem[{Scherrer {et~al.}(1995)Scherrer, Bogart, Bush, Hoeksema, Kosovichev,
  Schou, Rosenberg, Springer, Tarbell, Title, Wolfson, Zayer, \&
  Team}]{Scherrer1995}
Scherrer, P.~H., Bogart, R.~S., Bush, R.~I., {et~al.} 1995, Solar Physics, 162,
  129, \dodoi{10.1007/BF00733429}

\bibitem[{Schrijver(2007)}]{Schrijver2007}
Schrijver, C.~J. 2007, The Astrophysical Journal Letters, 655, L117.
\newblock \url{https://iopscience.iop.org/article/10.1086/511857}

\bibitem[{Selvaraju {et~al.}(2017)Selvaraju, Cogswell, Das, Vedantam, Parikh,
  \& Batra}]{Selvaraju2017}
Selvaraju, R.~R., Cogswell, M., Das, A., {et~al.} 2017, 2017 IEEE International
  Conference on Computer Vision (ICCV), 1, 618

\bibitem[{Shibata \& Magara(2011)}]{Shibata2011}
Shibata, K., \& Magara, T. 2011, Living Reviews in Solar Physics, 8, 6,
  \dodoi{10.12942/lrsp-2011-6}

\bibitem[{Simonyan {et~al.}(2013)Simonyan, Vedaldi, \&
  Zisserman}]{Simonyan2013}
Simonyan, K., Vedaldi, A., \& Zisserman, A. 2013, CoRR, abs/1312.6034, 1

\bibitem[{Stenflo(2013)}]{Stenflo2013}
Stenflo, J.~O. 2013, The Astronomy and Astrophysics Review, 21, 66,
  \dodoi{10.1007/s00159-013-0066-3}

\bibitem[{Su {et~al.}(2013)Su, Veronig, Holman, Dennis, Wang, Temmer, \&
  Gan}]{Su2013}
Su, Y., Veronig, A.~M., Holman, G.~D., {et~al.} 2013, Nature Physics, 9, 489 EP
  .
\newblock \url{http://dx.doi.org/10.1038/nphys2675}

\bibitem[{Sun {et~al.}(2022)Sun, Bobra, Wang, Wang, Sun, Gombosi, Chen, \&
  Hero}]{Sun_2022}
Sun, Z., Bobra, M.~G., Wang, X., {et~al.} 2022, The Astrophysical Journal, 931,
  163, \dodoi{10.3847/1538-4357/ac64a6}

\bibitem[{Sundararajan {et~al.}(2017)Sundararajan, Taly, \&
  Yan}]{pmlr-v70-sundararajan17a}
Sundararajan, M., Taly, A., \& Yan, Q. 2017, in Proceedings of Machine Learning
  Research, Vol.~70, Proceedings of the 34th International Conference on
  Machine Learning, ed. D.~Precup \& Y.~W. Teh (PMLR), 3319--3328.
\newblock \url{https://proceedings.mlr.press/v70/sundararajan17a.html}

\bibitem[{{Szegedy} {et~al.}(2015){Szegedy}, {Wei Liu}, {Yangqing Jia},
  {Sermanet}, {Reed}, {Anguelov}, {Erhan}, {Vanhoucke}, \&
  {Rabinovich}}]{GoogleNet2015}
{Szegedy}, C., {Wei Liu}, {Yangqing Jia}, {et~al.} 2015, in 2015 IEEE
  Conference on Computer Vision and Pattern Recognition (CVPR), 1--9,
  \dodoi{10.1109/CVPR.2015.7298594}

\bibitem[{Toriumi \& Wang(2019)}]{Toriumi2019}
Toriumi, S., \& Wang, H. 2019, Living Reviews in Solar Physics, 16, 3,
  \dodoi{10.1007/s41116-019-0019-7}

\bibitem[{Zeiler \& Fergus(2014)}]{Zeiler2014}
Zeiler, M., \& Fergus, R. 2014, in Lecture Notes in Computer Science (including
  subseries Lecture Notes in Artificial Intelligence and Lecture Notes in
  Bioinformatics), Vol. 8689 LNCS, Computer Vision, ECCV 2014 - 13th European
  Conference, Proceedings, part 1 edn. (Springer Verlag), 818--833,
  \dodoi{10.1007/978-3-319-10590-1_53}

\bibitem[{Zhang {et~al.}(2008)Zhang, Liu, \& Zhang}]{Zhang2008}
Zhang, Y., Liu, J., \& Zhang, H. 2008, Solar Physics, 247, 39,
  \dodoi{10.1007/s11207-007-9089-0}

\end{thebibliography}

\appendix

\section{Time Evolution of the CNN-estimated features on all ARs producing flares M5 or greater.}
\begin{figure*}[ht]
\centering
\subfloat{\includegraphics[width= 0.33\textwidth,trim={0.5cm 0.2cm 0.2cm 0.4cm},clip]{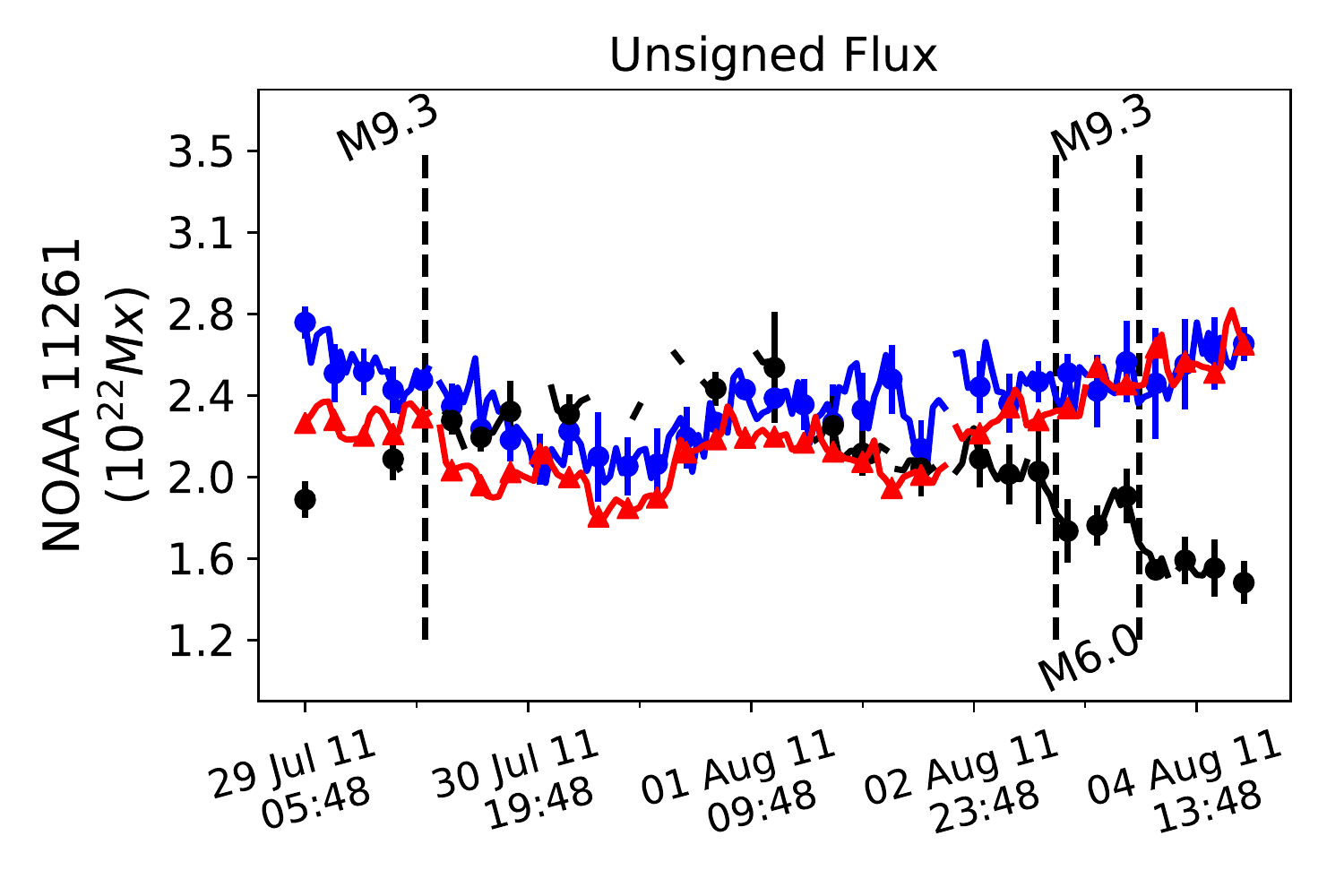}} 
\subfloat{\includegraphics[width= 0.33\textwidth,trim={0.5cm 0.2cm 0.2cm 0.4cm},clip]{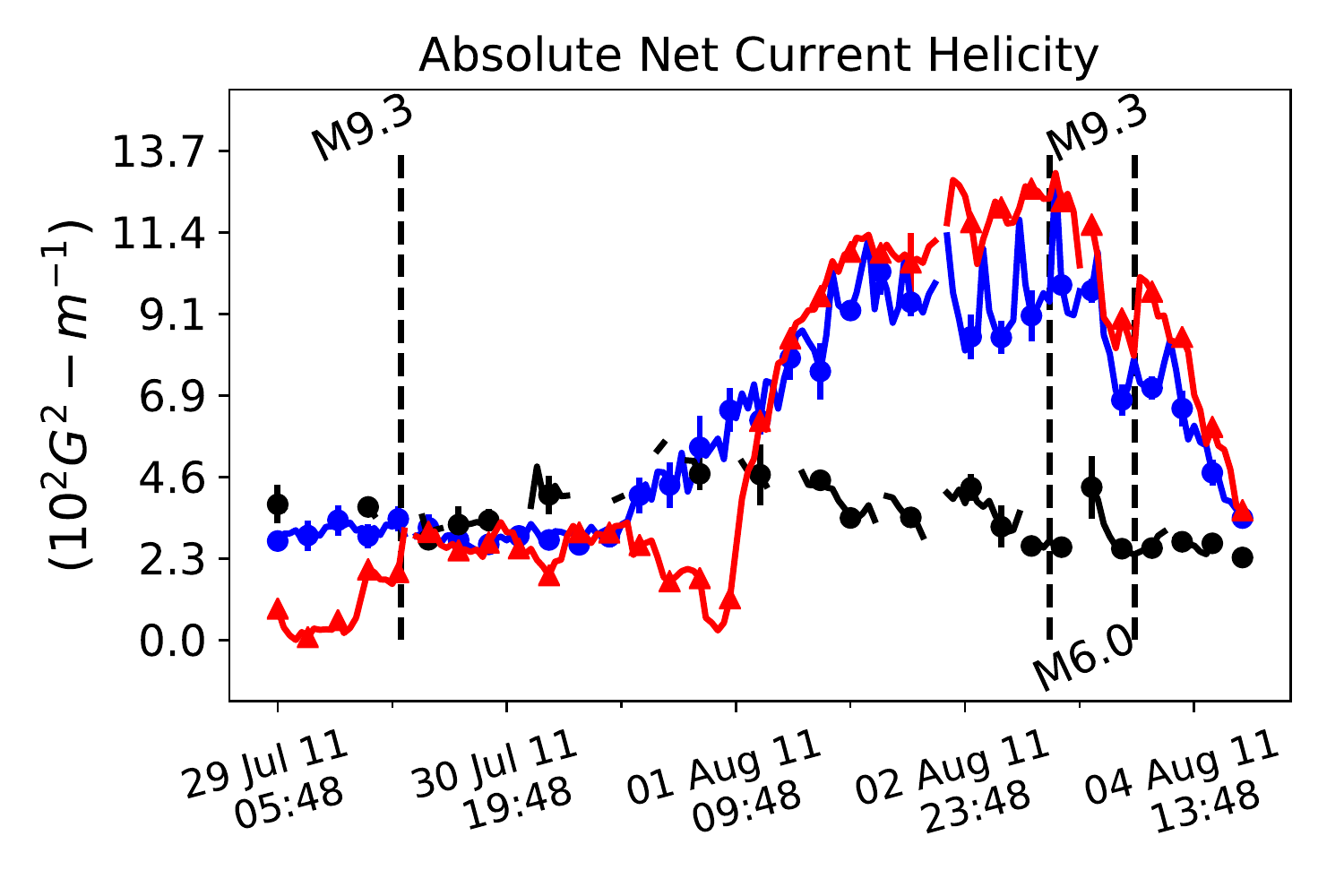}}
\subfloat{\includegraphics[width= 0.33\textwidth,trim={0.5cm 0.2cm 0.2cm 0.4cm},clip]{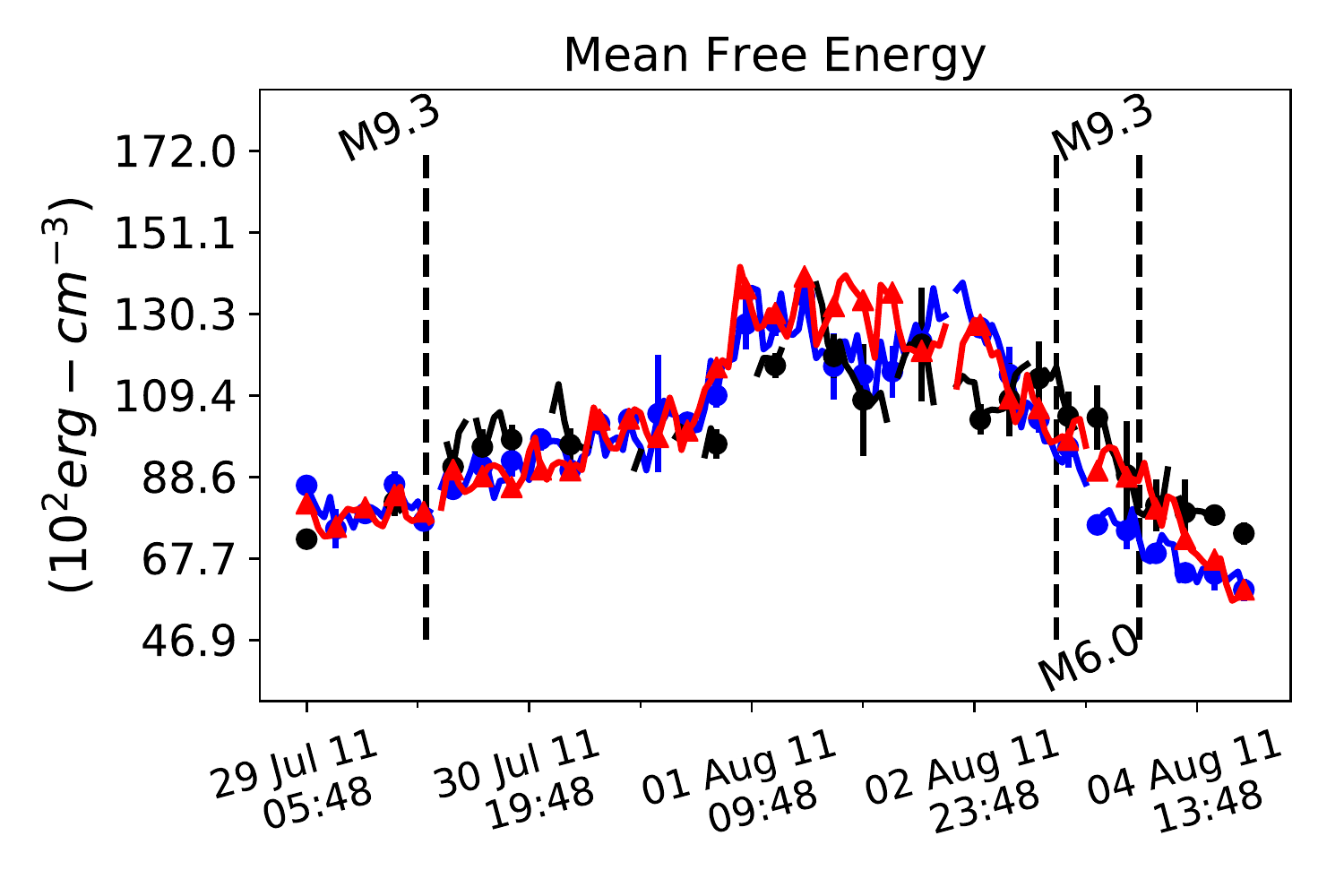}}\\
\caption{{\bf Time-evolution of the CNN-estimated vector-field features on flare-productive active regions.} Comparison of the CNN-estimated values, using HMI (blue) and GONG  (black) LOS magnetograms, of total unsigned flux, absolute net current helicity and mean free energy density with true values (red) calculated from HMI vector magnetograms are shown for HARP 750. Only observations within $\pm 45\degree$ are considered. 1-$\sigma$ error bars are shown. The gaps indicate missing observations. The complete figure set (28 images) for active regions that produced at least one M5 or greater flare is available in the online journal.}
\label{fig:allTE}
\end{figure*}

\section{Comparison of the CNN-estimated features during the 2003 Halloween storms}
\begin{figure*}[ht]
\centering
\subfloat{\includegraphics[width= 0.33\textwidth,trim={0.5cm 0.2cm 0.2cm 0.4cm},clip]{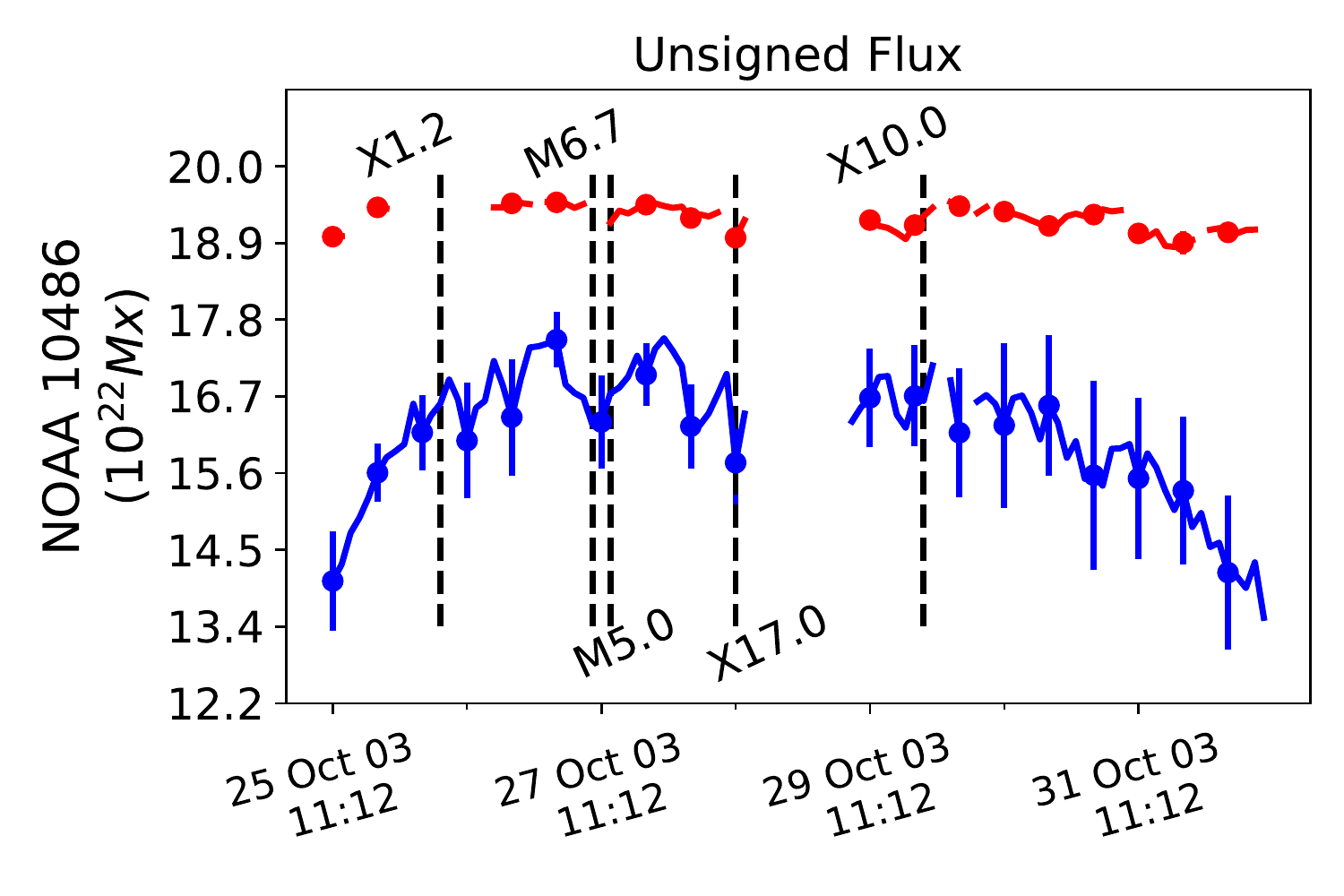}} 
\subfloat{\includegraphics[width= 0.33\textwidth,trim={0.5cm 0.2cm 0.2cm 0.4cm},clip]{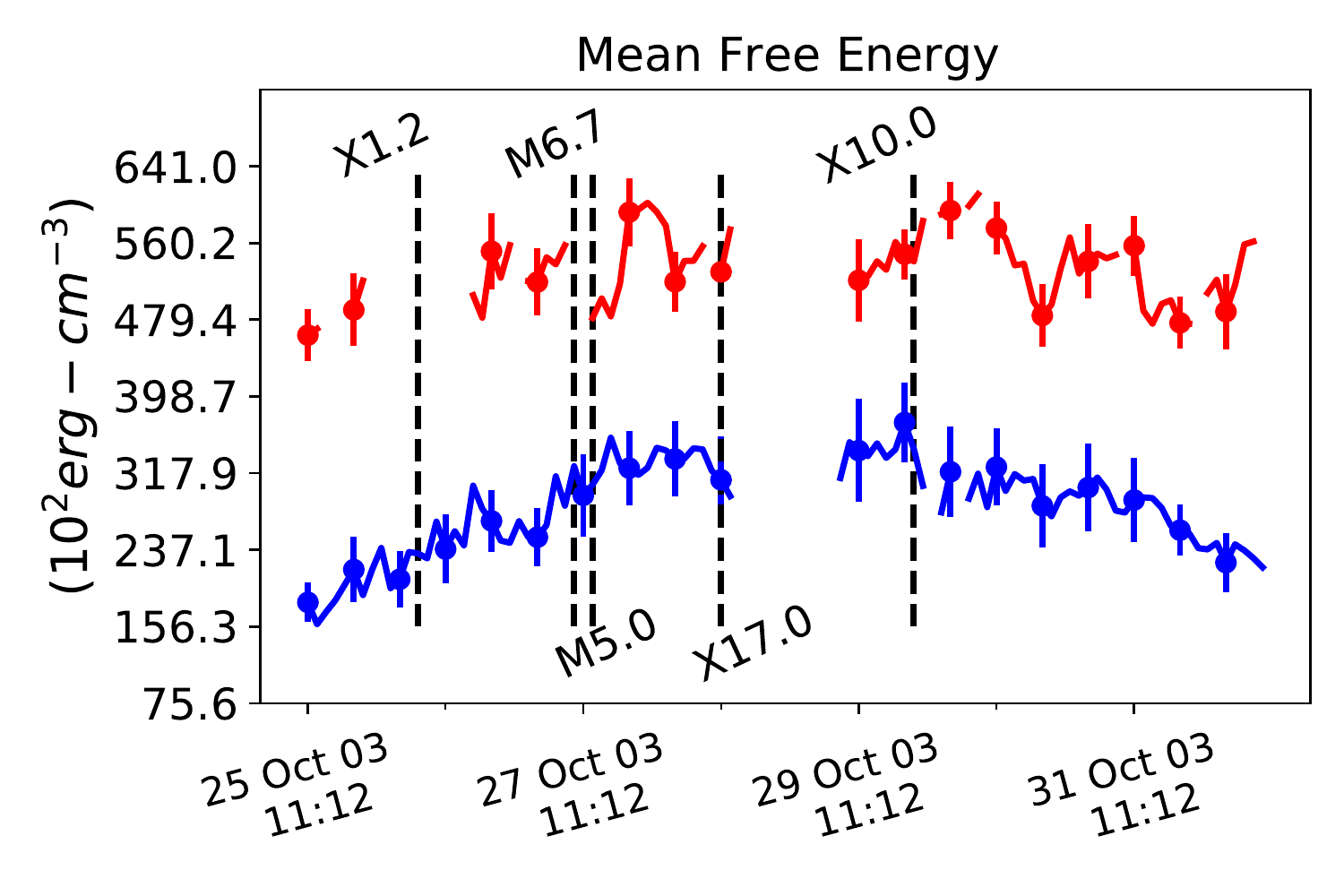}}
\subfloat{\includegraphics[width= 0.33\textwidth,trim={0.5cm 0.2cm 0.2cm 0.4cm},clip]{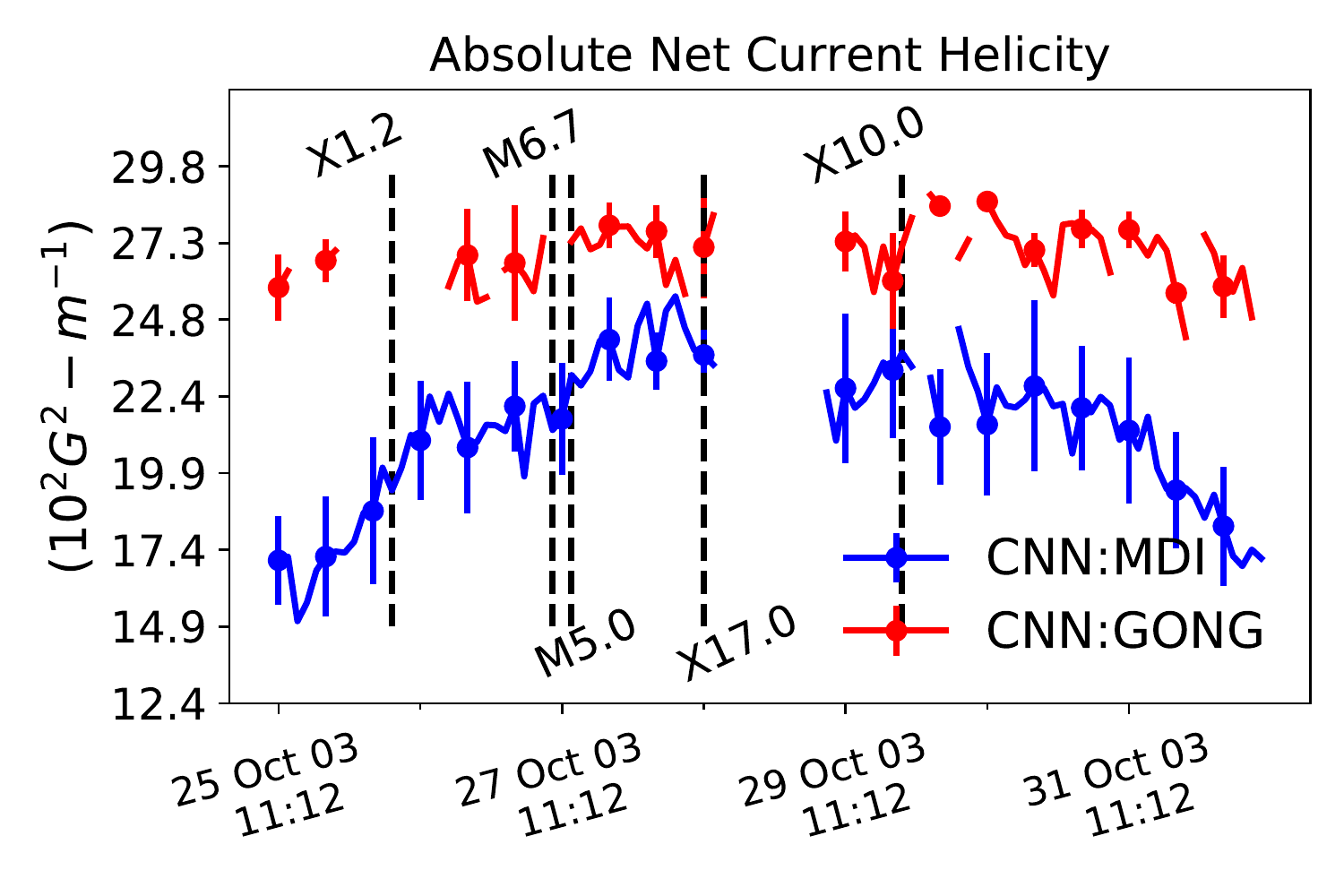}}\\
\caption{{\bf A comparison of the time-evolution of the CNN-estimated vector-field features during the 2003 Halloween storms using MDI and GONG LOS magnetograms.} The CNN trained with HMI magnetograms is used for the estimation from MDI magnetograms whereas the CNN trained with GONG magnetograms is used for the estimation from GONG magnetograms. Note that the HMI observations are not available before 2010. The CNN:GONG feature values are generally high compared to the CNN:MDI, showing little variation throughout the storms. The CNN:MDI features appear to capture the variation of these features during the storms expected from the theoretical modelling e.g. \citep{Kazachenko_2010}. The 1-$\sigma$ errors are shown. The gaps indicate missing observations. The legend in the rightmost panel applies to all panels. }
\label{fig:allHallow}
\end{figure*}

\section{MDI Correlations}
\begin{table*}[t]
\centering
\begin{tabular}{lcc}
\hline
\hline
 SHARPs Features & Pearson correlation & Spearman correlation \\
\hline
Total unsigned flux               & 82.49 $\pm$ 21.01 & 68.09 $\pm$ 37.73  \\
Area                              & 85.55 $\pm$ 23.82 & 74.59 $\pm$ 17.72  \\ 
Total unsigned vertical current   & 74.83 $\pm$ 45.39 & 64.60 $\pm$ 46.31  \\
Total unsigned current helicity   & 75.22 $\pm$ 45.28 & 65.80 $\pm$ 47.14  \\
Total free energy density         & 84.18 $\pm$ 12.15 & 76.40 $\pm$ 19.94  \\ 
Total Lorentz force               & 89.13 $\pm$ 11.33 & 75.37 $\pm$ 29.10  \\
\hline
Absolute net current helicity     & 51.62 $\pm$ 27.28 &  48.97 $\pm$ 23.21 \\
Sum of net current per polarity   & 42.84 $\pm$ 35.65 &  38.69 $\pm$ 32.27 \\
\hline
Mean free energy density          & 92.60 $\pm$ 04.02 & 89.26 $\pm$ 07.80  \\ 
Area with shear $>45\degree$      & 91.78 $\pm$ 03.32 & 89.63 $\pm$ 03.32  \\ 
\hline
Flux near polarity inversion line & 62.64 $\pm$ 14.63 & 59.09 $\pm$ 21.23  \\ 
\hline
\hline
\end{tabular}
\caption{{\bf Pearson and Spearman correlations between the CNN-estimated vector-field features SHARPs using MDI line-of-sight magnetograms and their true values.} The SHARPs features are estimated using the CNN trained with the HMI line-of-sight magnetograms. The AR patches of MDI line-of-sight magnetograms are taken from the publicly available data product Space-Weather MDI Active Region Patches (SMARPs) \citep{Bobra_2021}. SMARPs and SHARPs data overlap between 1 May 2010 and 28 October 2010 \citep{Bobra_2021}.}
\label{tab:dataNcorrMDI}
\end{table*}
\newpage
\begin{figure*}
\centering
\includegraphics[width= 0.49\textwidth,trim={1.75cm 1.2cm 2.0cm 0.7cm},clip]{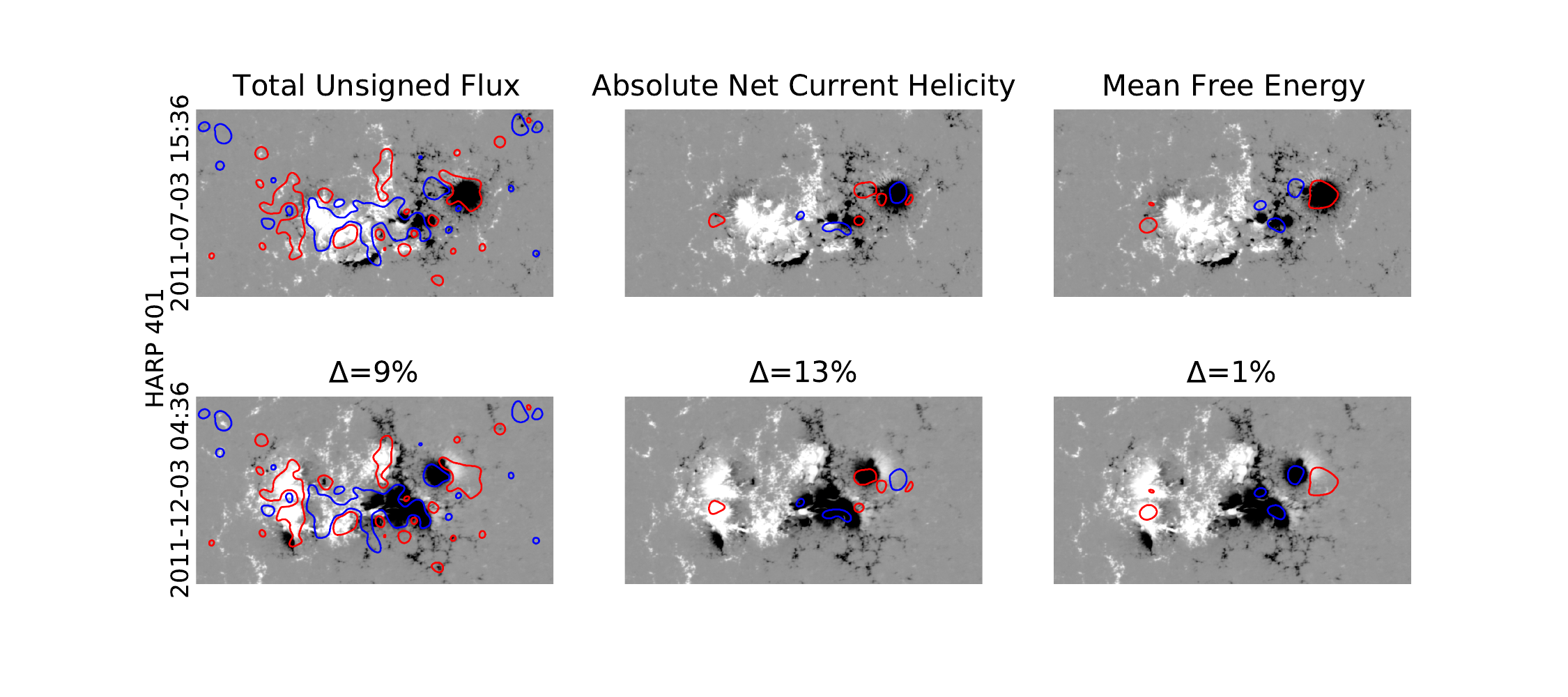}
\includegraphics[width= 0.49\textwidth,trim={1.75cm 1.2cm 2.0cm 0.7cm},clip]{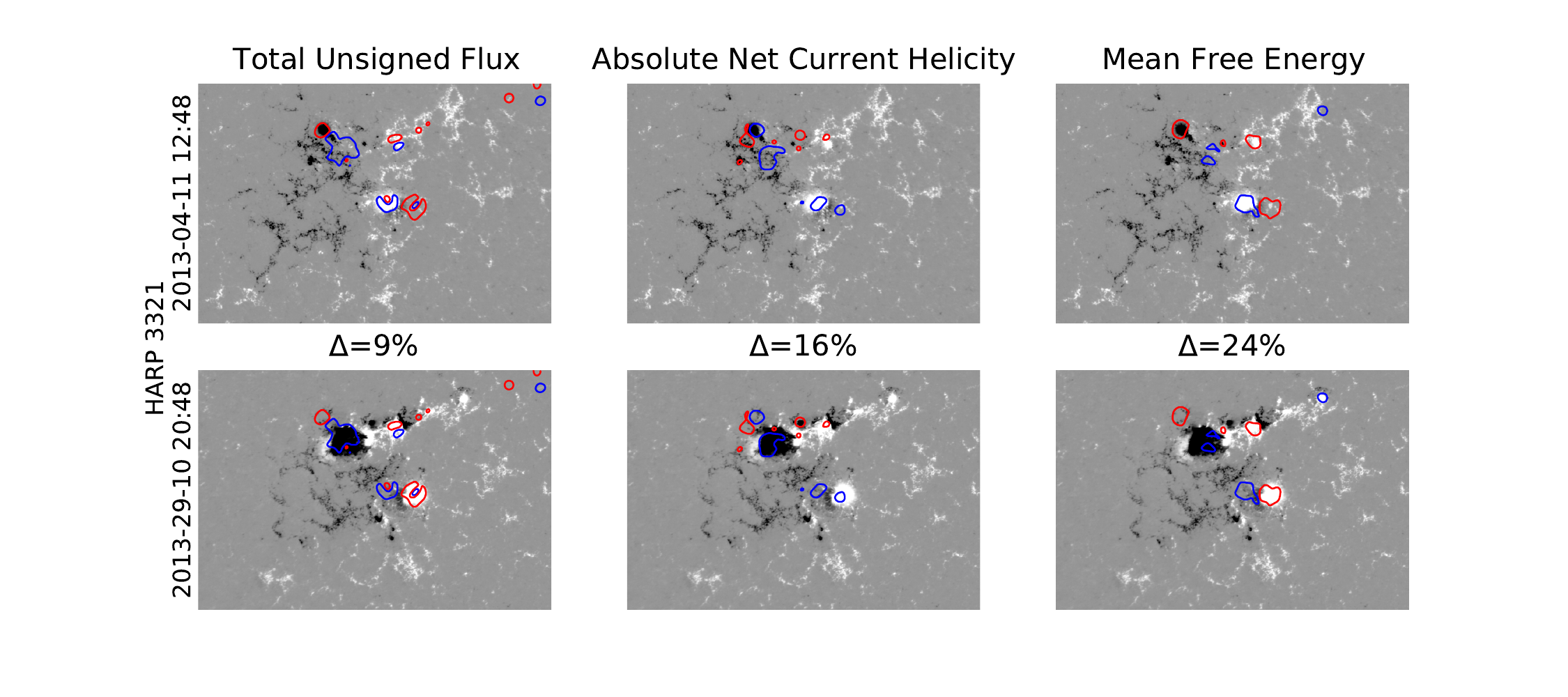}\\
\vspace{\baselineskip}
\includegraphics[width= 0.49\textwidth,trim={1.75cm 1.2cm 2.0cm 0.7cm},clip]{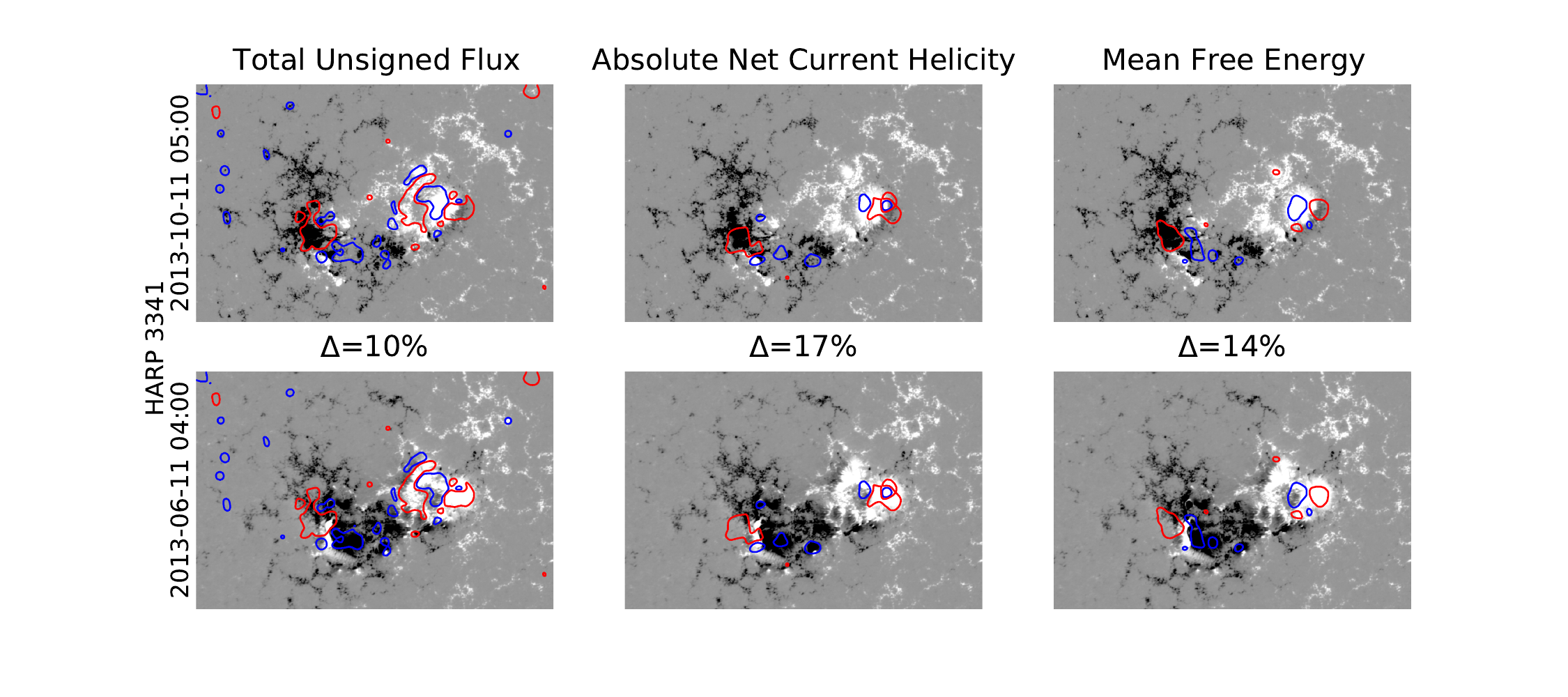}
\includegraphics[width= 0.49\textwidth,trim={1.75cm 1.2cm 2.0cm 0.7cm},clip]{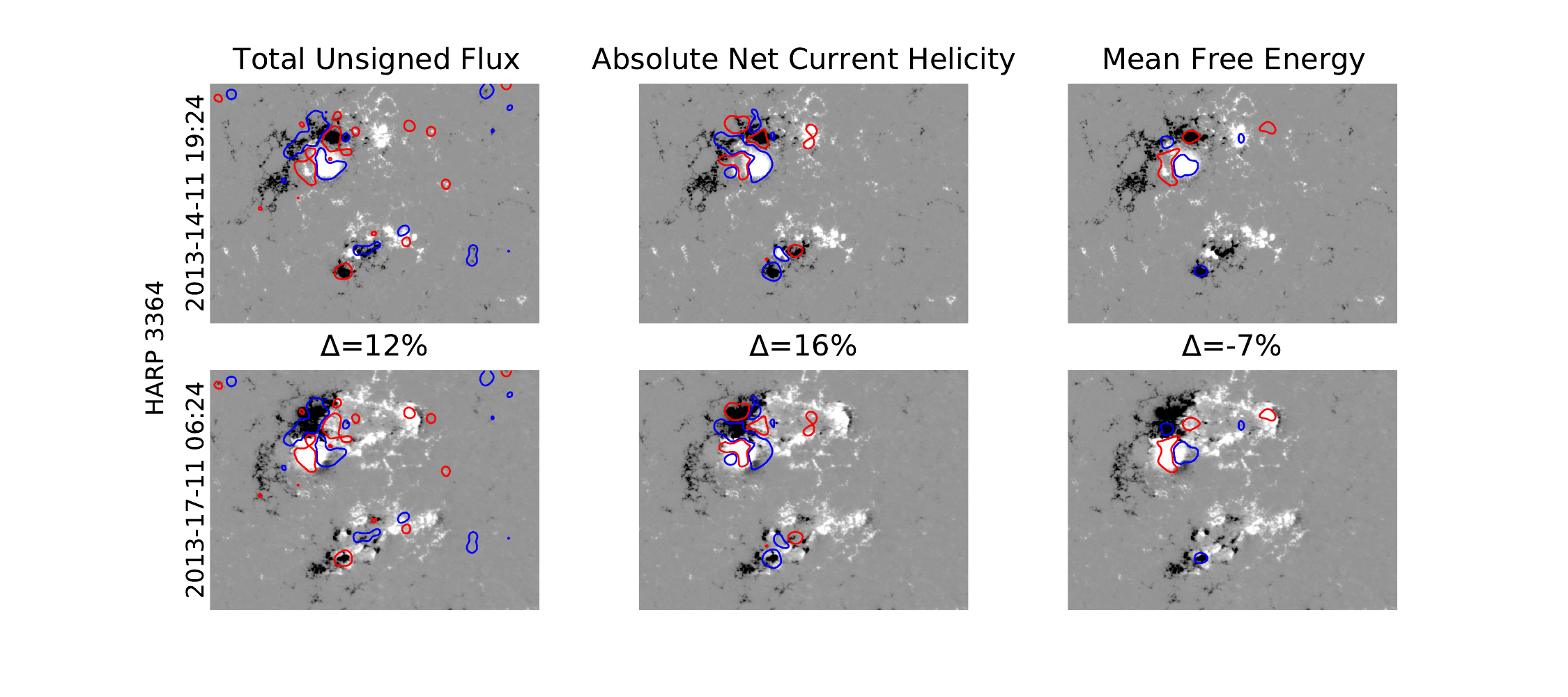}\\
\vspace{\baselineskip}
\includegraphics[width= 0.49\textwidth,trim={1.75cm 1.2cm 2.0cm 0.7cm},clip]{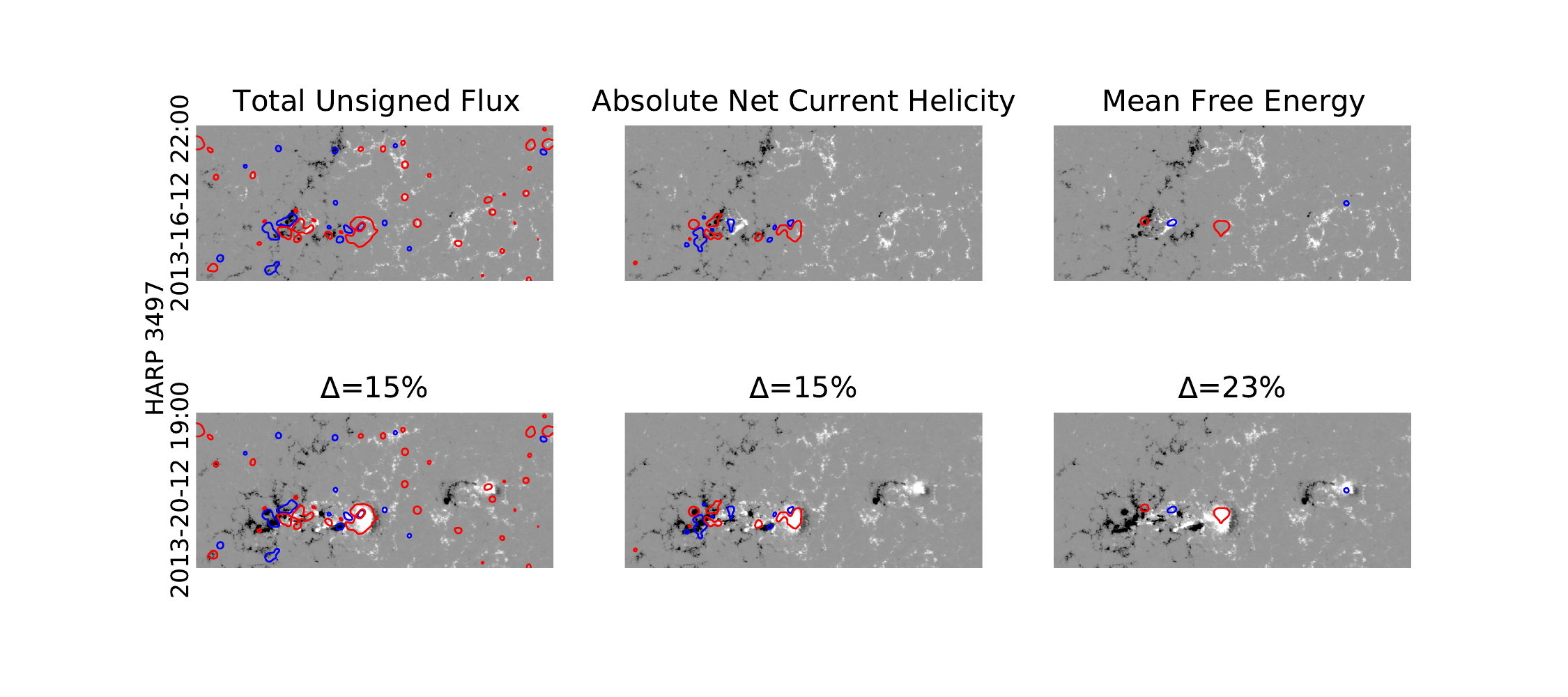}
\includegraphics[width= 0.49\textwidth,trim={1.75cm 1.2cm 2.0cm 0.7cm},clip]{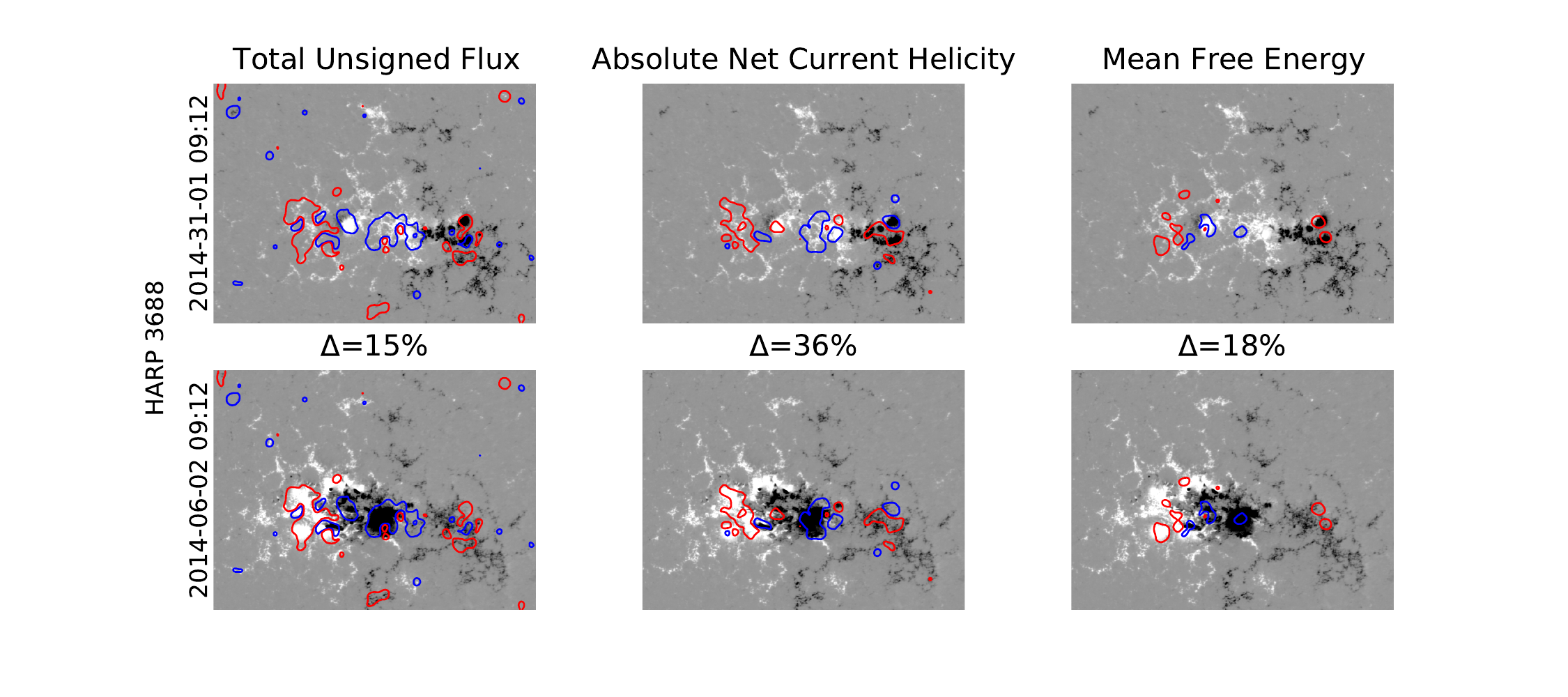}\\
\includegraphics[width= 0.65\textwidth,trim={1.75cm 1.2cm 2.0cm 0.7cm},clip]{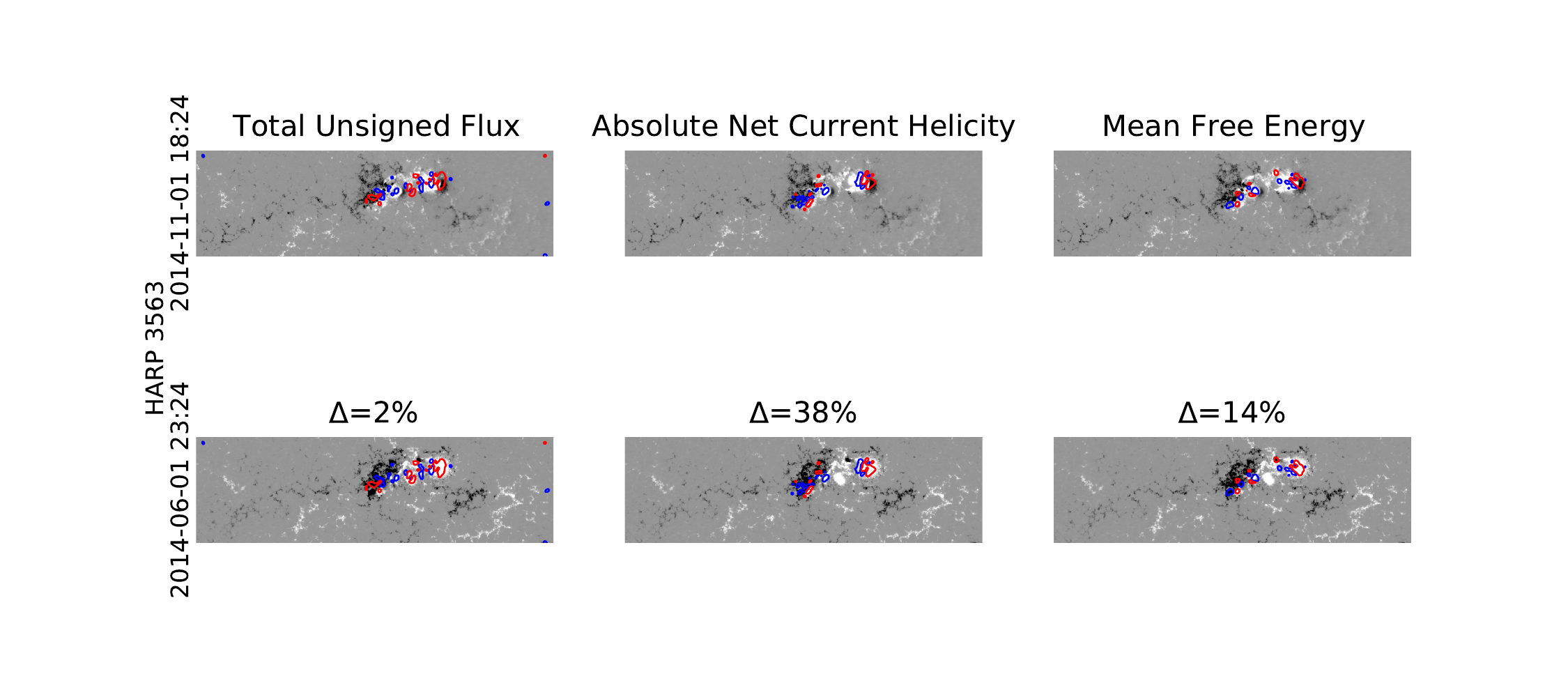}\\
\includegraphics[width= 0.65\textwidth,trim={1.75cm 1.2cm 2.0cm 0.7cm},clip]{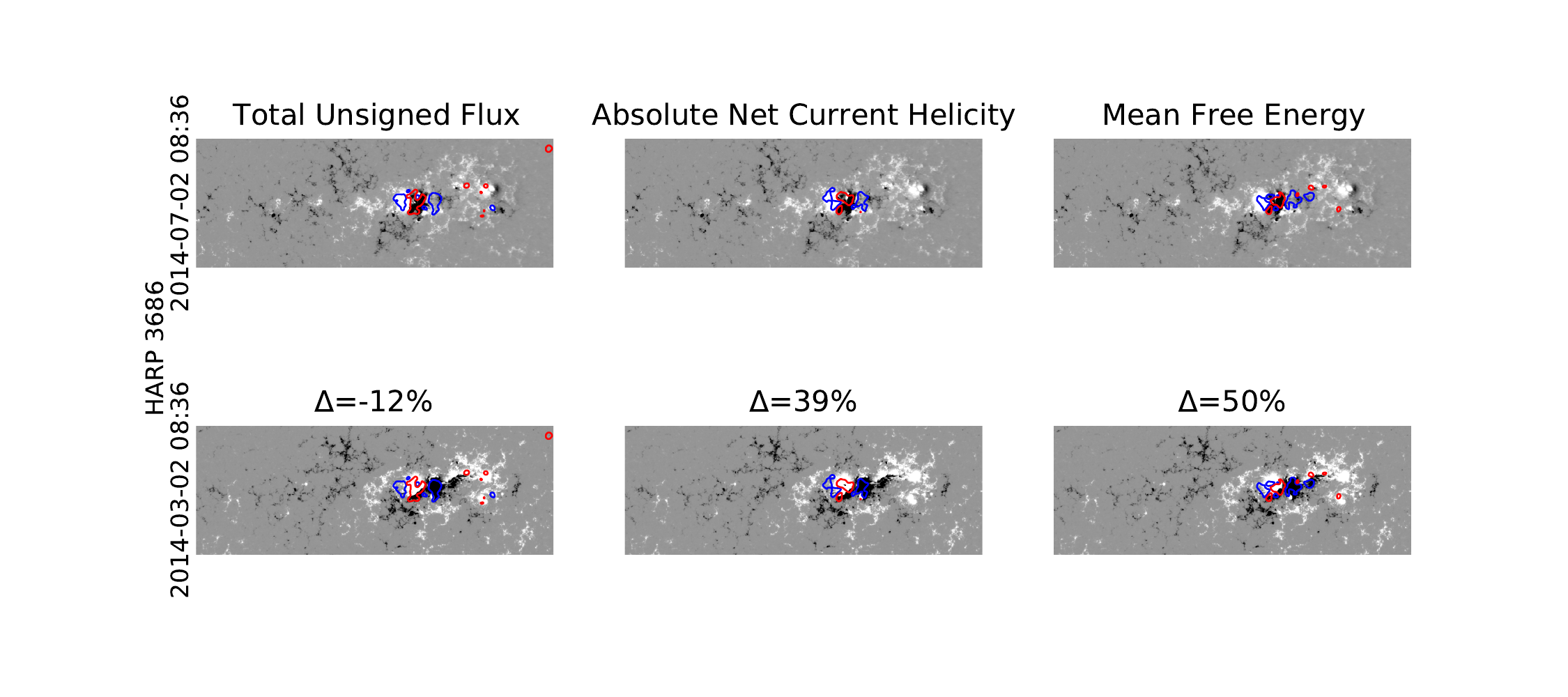}
\caption{{\bf Additional examples of Integrated gradient (IG) attribution maps (Figure~\ref{fig:IG}).}} 
\label{fig:IGApp}
\end{figure*}
\end{document}